\documentclass[a4paper,onecolumn,11pt]{quantumarticle}
\pdfoutput=1
\usepackage[utf8]{inputenc}
\usepackage[english]{babel}
\usepackage[T1]{fontenc}
\usepackage{amsmath}
\usepackage{amsfonts}
\usepackage{graphicx}
\usepackage{float}
\usepackage{placeins}
\usepackage{hyperref}
\usepackage{physics}
\usepackage{tikz}
\usepackage[numbers, sort&compress]{natbib}

\usepackage{algorithm}
\usepackage{algpseudocode}

\usepackage{listings}
\definecolor{codegreen}{rgb}{0,0.6,0}
\definecolor{codegray}{rgb}{0.5,0.5,0.5}
\definecolor{codepurple}{rgb}{0.58,0,0.82}
\definecolor{codered}{rgb}{0.796,0.235,0.2}
\definecolor{backcolour}{rgb}{0.95,0.95,0.92}
\definecolor{dodgerblue4}{rgb}{0.06, 0.31, 0.55}

\lstdefinelanguage{Julia}{
    keywords=[1]{abstract, break, catch, const, continue, do, else, elseif, begin, end, export, false, for, function, if, import, let, local, macro, module, quote, return, struct, true, try, using, while, where},
    keywords=[2]{Bool, Char, Dict, Float64, Int, String, Array, AbstractArray, Vector, Set, Tuple, NamedTuple, QuantumObject, QuantumObjectEvolution},
    keywords=[3]{dummy_start, versioninfo, Qobj, QobjEvo, tensor, basis, fock, destroy, qeye, sigmax, sigmay, sigmaz, sigmam, sigmap, range, sesolve, mesolve, mcsolve, ssesolve, smesolve, sqrt, sin, cos, steadystate, steadystate_fourier, dfd_mesolve, dsf_mesolve, coherent, sum, size, dropdims, H_dsf, c_ops_dsf, e_ops_dsf, cu, CuVector, wigner, normalize, addprocs, SlurmManager, set_num_threads, Lattice, mapreduce, multisite_operator, DissipativeIsing, Val, ntuple, EnsembleSplitThreads, rmprocs, workers, gradient, my_f_mesolve, real, expect, BacksolveAdjoint, EnzymeVJP, liouvillian, exp, conj, coef_a, coef_ac, dummy_end},
    sensitive=true,  % Julia is case-sensitive.
    morecomment=[l]\#,%
    morecomment=[n]{\#=}{=\#},%
    morestring=[s]{"}{"},%
    literate={
        {\\}{{{\color{codered}\lstum@backslash}}}{1} {\{}{{{\color{codered}\{}}}{1}
        {\}}{{{\color{codered}\}}}}{1} % {!}{{{\color{codered}!}}}{1}
        {\%}{{{\color{codered}\%}}}{1} {&}{{{\color{codered}\&}}}{1}
        {+}{{{\color{codered}+}}}{1}
        {*}{{{\color{codered}*}}}{1}
        {/}{{{\color{codered}/}}}{1}
        {'}{{{\color{codered}\textquotesingle}}}{1}
        {=}{{{\color{codered}=}}}{1}
        {<}{{{\color{codered}<}}}{1} 
        {=>}{{{\color{codered}=>}}}{1} {|>}{{{\color{codered}|>}}}{1}
        {==}{{{\color{codered}==}}}{1}
        {>}{{{\color{codered}>}}}{1} {?}{{{\color{codered}?}}}{1}
        {julia>}{{{\color{codered}julia>}}}{1}
        {^}{{{\color{codered}\textasciicircum}}}{1} {|}{{{\color{codered}|}}}{1}
        {~}{{{\color{codered}\textasciitilde{}}}}{1}
        {×}{{{\color{codered}$\times$}}}{1} {⋅}{{{\color{codered}$\cdot$}}}{1}
        {π}{{$\pi$}}1
        {ω}{{$\omega$}}1
        {α}{{$\alpha$}}1
        {β}{{$\beta$}}1
        {σ}{{$\sigma$}}1
        {ψ}{{$\psi$}}1
        {κ}{{$\kappa$}}1
        {γ}{{$\gamma$}}1
        {κ}{{$\kappa$}}1
        {Δ}{{$\Delta$}}1
        {φ}{{$\phi$}}1
        {∘}{{$\circ$}}1
    }
}

\lstdefinestyle{mainstyle}{
    language=Julia,
    backgroundcolor=\color{backcolour},   
    commentstyle=\color{gray},    % Style for comments.
    keywordstyle=\color{blue}\bfseries,  % Style for Group 1 keywords.
    keywordstyle=[2]\color{teal},  % Style for Group 2 keywords.
    keywordstyle=[3]\color{dodgerblue4},  % Style for Group 3 keywords.
    stringstyle=\color{codepurple},
    basicstyle=\ttfamily\footnotesize,
    columns=fullflexible,
    upquote=true,
    breakatwhitespace=false,         
    breaklines=true,                 
    captionpos=b,                    
    keepspaces=true,                 
    showspaces=false,                
    showstringspaces=false,
    showtabs=false,                  
    tabsize=2,
}

\lstdefinestyle{tablestyle}{
    language=Julia,
    backgroundcolor=\color{backcolour},
    keywordstyle=[2]\color{black},  % Style for Group 2 keywords.
    keywordstyle=[3]\color{black},  % Style for Group 3 keywords.
    basicstyle=\ttfamily\footnotesize,
    columns=fullflexible,
    upquote=true,
    breakatwhitespace=false,         
    breaklines=true,                 
    captionpos=b,                    
    keepspaces=true,                 
    numbers=left,                    
    numbersep=5pt,                  
    showspaces=false,                
    showstringspaces=false,
    showtabs=false,                  
    tabsize=2,
}

\newcommand{\code}[1]{%
  \begingroup
  \setlength{\fboxsep}{1pt} % Reduce padding inside the box
  \colorbox{backcolour}{\lstinline[basicstyle=\ttfamily]{#1}}%
  \endgroup
}

\begin{document}

% \title{QuantumFCS.jl: A Numerical Tool for Full-Counting Statistics}
% \title{QuantumFCS.jl: Open-Source Numerical Tool for Full-Counting Statistics}
\title{QuantumFCS.jl: Efficient Full-Counting Statistics for Open Quantum Systems}

\author{Aaron Daniel}
\email{aaron.daniel@unibas.ch}
%\homepage{http://quantum-journal.org}
\orcid{0000-0003-1123-178X}
\affiliation{
Department of Physics and Swiss Nanoscience Institute, University of Basel, Klingelbergstrasse 82, CH-4056 Basel, Switzerland\
}

\author{Marcelo Janovitch}
\email{m.janovitch@unibas.ch}
\affiliation{
Department of Physics and Swiss Nanoscience Institute, University of Basel, Klingelbergstrasse 82, CH-4056 Basel, Switzerland\
}
\orcid{0000-0003-3614-0622}

\maketitle

\begin{abstract}
Full-counting statistics (FCS) provides a systematic framework for characterising current fluctuations in quantum transport, quantum optics, and open quantum systems. We introduce \texttt{QuantumFCS.jl}, an open-source Julia package for efficient and flexible numerical FCS calculations. The package defines currents through monitored jump operators and weights, allowing particle, electric, and heat currents to be treated within the same workflow. It implements a recursive cumulant algorithm with dense, sparse, and iterative solver backends, making it suitable for models that are challenging for direct dense methods. 
% \revblue{For dimensionless event-counting observables, the same interface also returns factorial cumulants, which are useful diagnostics for deviations from independent Poissonian counting.}
We demonstrate the package on photon counting in a driven-dissipative Jaynes-Cummings system near the blockade-breakdown phase transition and on heat-current fluctuations in a non-linear circuit-QED heat engine. These examples show how higher cumulants reveal intermittency between bright and dim emission, antibunching, and provide probes of thermodynamic uncertainty relations. Benchmarks against an existing FCS implementation show substantial speed-ups.
\end{abstract}

\tableofcontents

\section{Introduction}
\label{sec:introduction}
Full counting statistics (FCS) has become a central framework for characterising fluctuations in quantum transport and open quantum systems, extending the description of transport beyond average currents and simple two-point correlation functions~\cite{Levitov1993, Levitov1996, Nazarov2002, Nazarov2003, Belzig2005, Kaiser2007, Flindt2008, Flindt2010, Ubbelohde2011, Flindt2013, Esposito2014, Potts2016, Brange2019, Landi2024, potts2026quantumthermodynamics, Li2026, Menczel2026}. This framework forms part of the modern description of continuously monitored open quantum systems, where multi-time correlation functions, quantum trajectories, and FCS provide complementary descriptions of the stochastic measurement records~\cite{PhysRevLett.68.580,Landi2024}.

In the context of mesoscopic conductors, FCS reveals the role of Pauli exclusion in suppressing current fluctuations in non-interacting systems~\cite{Levitov1993, Levitov1996} and is powerful for probing interaction effects such as the Coulomb blockade~\cite{Flindt2010,PhysRevB.67.085316, Kambly2011Factorial, Stegmann2015GeneralizedFactorial, Komijani2013HoleTransfer}. 
%\revblue{For discrete electron-counting variables, factorial cumulants provide a complementary representation in which interaction-induced correlations can be particularly transparent~\cite{Kambly2011Factorial,Stegmann2015GeneralizedFactorial,Komijani2013HoleTransfer}.} 
Beyond electronic transport, FCS has also emerged as an important tool in quantum optics and quantum thermodynamics. Photon counting statistics has been used to analyse bosonic bunching and antibunching~\cite{Brange2019} and, more broadly, counting statistics provides a natural framework for studying fluctuation relations and heat-current fluctuations in quantum systems~\cite{Esposito2014, potts2026quantumthermodynamics, Brange2019}.
Furthermore, FCS can be used to probe dynamical signatures of non-classicality~\cite{Potts2016} and phase transitions~\cite{Flindt2013, Google2024, Valli2025, Menczel2026}. These developments establish FCS as a central tool for diagnosing the role of noise, interactions, and non-classical phenomena in open quantum systems.

Despite the importance and maturity of FCS, to the best of our knowledge there is currently no publicly available, open-source software package dedicated to the numerical computation of FCS to arbitrary cumulant order. The only comparable public tool we are aware of, the Mathematica package \texttt{Melt}~\cite{Melt}, provides the first two current cumulants and serves as the benchmark comparator in Sec.~\ref{sec:benchmarking}.

Here, we introduce \texttt{QuantumFCS.jl}~\cite{QuantumFCS}, a numerical package written in the Julia programming language for computing the FCS of quantum systems described by Lindblad master equations. The package is centred around a function that computes cumulants of arbitrary order using the recursive scheme introduced in Refs.~\cite{Flindt2008, Flindt2010}. Combined with the high-performance capabilities of Julia and its growing open-source scientific-computing ecosystem \cite{doi:10.1137/141000671}, our implementation aims to lower the barrier to entry for FCS analysis. Three features distinguish it from existing practice: current cumulants of arbitrary order are generated directly from the recursive scheme, avoiding numerical differentiation of the tilted-Liouvillian eigenvalue; dense, sparse, and iterative solver backends are exposed through a single interface; and an iterative Drazin backend based on incomplete-LU-preconditioned GMRES makes full-counting statistics of large sparse Liouvillians---with vectorised dimension exceeding $10^6$---tractable on a tabletop machine. As shown in Sec.~\ref{sec:benchmarking}, the standard LU-based dense and sparse backends also yield substantial speed-ups over the only existing comparator.

In this manuscript we illustrate the package on three representative applications. We first use a minimal single-level quantum dot to show the basic workflow and to connect with standard electron-counting settings~\cite{PhysRevLett.96.076605}. We then consider photon emission from a driven-dissipative Jaynes-Cummings system close to the blockade-breakdown crossover, where higher cumulants reveal strongly super-Poissonian statistics and intermittency between photon-blockade regimes (dim) and bursts of photons (bright)~\cite{Carmichael2015BreakdownPhotonBlockade}. Finally, we apply the package to a circuit-QED heat engine based on photon-assisted Cooper-pair tunnelling~\cite{HoferSouquetClerk2016,Kerremans2022,Janovitch2023}, where FCS provides direct access to heat-current fluctuations, antibunching, and thermodynamic uncertainty relations for heat currents~\cite{Barato2015, Pietzonka2016, Pietzonka2017, Gingrich2016, Horowitz2020, AgarwallaSegal2018, Liu2019, Falasco2020, Prech2023, Ptaszyski2023, AlmanzaMarreroManzano2025}.

The paper is organised as follows. In Sec.~\ref{sec:theory}, we introduce the Lindblad-FCS setting and the recursive cumulant algorithm. In Sec.~\ref{sec:usage}, we describe the package interface, demonstrate the basic workflow on a minimal quantum-dot example, and then introduce the sparse iterative solver backend. Sections~\ref{sec:ddjc} and~\ref{sec:circuit-qhe} present the Jaynes-Cummings and heat-engine applications, respectively. We then benchmark the implementation in Sec.~\ref{sec:benchmarking} and close with a summary and discussion of possible extensions in Sec.~\ref{sec:summary}.

\section{Setup and theoretical background}
\label{sec:theory}

\paragraph{Lindblad dynamics.}
Our implementation focuses on quantum systems described by time-independent Lindblad master equations \cite{Lindblad1976},
\begin{align}
    \dv{t}\rho &= -i[H,\rho] + \sum_{k=1}^r \mathcal{D}[J_k]\rho \equiv \mathcal{L}\rho ,\label{eq:LME}\\
    J_k &= \sqrt{\Gamma_k} L_k .
\end{align}
Here, $\rho$ is the system density matrix, $H$ is the system Hamiltonian, and $J_k$ are the jump operators. If the operators $L_k$ are dimensionless, the parameters $\Gamma_k$ are rates, $[\Gamma_k]=\mathrm{time}^{-1}$, and the jump operators have dimensions $[J_k]=\mathrm{time}^{-1/2}$. We use the Lindblad dissipator
\begin{align}
    \mathcal{D}[A]\rho = A\rho A^\dagger - \frac{1}{2}\{A^\dagger A,\rho\},
\end{align}
and set $\hbar=1$ throughout. The Liouvillian superoperator $\mathcal{L}$ is fully specified by the pair $(H,\mathbf{J})$, where $\mathbf{J}=(J_1,\ldots,J_r)$.

Equation~\eqref{eq:LME} describes Markovian open-system dynamics and is usually obtained under the standard weak-coupling and Born-Markov approximations \cite{BreuerMasterEquations}. The bath correlations are then local on the timescale resolved by the system dynamics, and the generator is time independent in the chosen frame. Coherently driven systems can therefore be treated after moving to a suitable rotating frame. For the recursive construction below, we assume that $\mathcal{L}$ has a unique stationary state $\rho_{\mathrm{ss}}$ and that the corresponding zero eigenvalue is isolated. Extensions to non-Markovian environments and genuinely time-dependent generators are discussed in Sec.~\ref{sec:summary}.

\paragraph{Currents from monitored jumps.}
The jump operators in Eq.~\eqref{eq:LME} can describe different exchange processes with one or several reservoirs. To define a current, we choose a list of monitored jumps
\begin{align}
    \mathbf{M} = (M_1,\ldots,M_p), \qquad M_l \in \{J_1,\ldots,J_r\},
\end{align}
and a corresponding list of weights $\boldsymbol{\nu}=(\nu_1,\ldots,\nu_p)$. In the package interface, these two objects are passed as \code{mJ} and \code{nu}. The weights fix both the orientation and the physical units of the counted current. They may be dimensionless for particle or photon currents, or dimensional for energy and heat currents. Together with the system data $(H,\mathbf{J})$ or an already assembled Liouvillian $\mathcal{L}$, these monitored jumps and weights are the core inputs of a \code{LindbladFCS} problem.

As a simple example, consider a bosonic mode coupled to two dissipative channels,
\begin{align}
    H &= \Omega a^\dagger a, \label{eq:ham-motivation}\\
    J_1 &= \sqrt{\kappa_A(\bar n_A+1)}\,a, \qquad
    J_2 = \sqrt{\kappa_A \bar n_A}\,a^\dagger, \label{eq:A-jumps-motivation}\\
    J_3 &= \sqrt{\kappa_B}\,a^2. \label{eq:B-jumps-motivation}
\end{align}
Here, $J_1$ and $J_2$ describe single-photon emission to and absorption from bath $A$, whose Bose-Einstein occupation is denoted by $\bar n_A$. The operator $J_3$ describes a two-photon loss process into channel $B$. Depending on the question of interest, different choices of monitored jumps and weights are useful. 

\begin{table}[t]
    \centering
    \begin{tabular}{c|c|c}
       $I(t)$ & $\mathbf{M}$ & $\boldsymbol{\nu}$\\
         \hline
        Net photon current from bath $A$ into the system & $(J_1,J_2)$ & $(-1,+1)$\\
        Heat current from bath $A$ into the system & $(J_1,J_2)$ & $(-\Omega,+\Omega)$\\
        Net photon current out of the system & $(J_1,J_3)$ & $(+1,+2)$
    \end{tabular}
    \caption{Example choices of monitored jumps and weights for the model defined in Eqs.~(\ref{eq:ham-motivation}--\ref{eq:B-jumps-motivation}). The sign convention is fixed by the weights. For the first two rows, positive current corresponds to photons or energy entering the system from bath $A$. In the last row, positive current corresponds to photons leaving the system.}
    \label{tab:jumps-and-weights}
\end{table}

The examples in Table~\ref{tab:jumps-and-weights} illustrate that the same Liouvillian can be used to study different stochastic currents. For the net photon current associated with bath $A$, emission and absorption events are counted with opposite signs. For the corresponding heat current, the same jump events are weighted by the energy quantum $\Omega$. For the total photon current out of the system, the two-photon loss channel contributes twice the weight of a single-photon emission event.

\paragraph{Tilted Liouvillian.}
Let $N_l(t)$ denote the number of jump events associated with the monitored jump $M_l$ during the time interval $[0,t]$. The net number of monitored jump events,
\begin{align}
    N(t) = \sum_{l=1}^p \nu_l N_l(t) = \int_0^t \dd t'\, I(t') ,
\end{align}
implicitly defines the stochastic current, $I(t)$, since its time integral corresponds to $N(t)$. To generate the statistics of $N(t)$, we introduce the counting-field-resolved density matrix
\begin{align}\label{eq:dressed-state}
    \rho_\chi(t) = \sum_{n\in\mathcal{N}} e^{i\chi n}\rho(n;t),
\end{align}
where $\mathcal{N}$ denotes the set of possible values of the integrated current. The operator $\rho(n;t)$ is the unnormalised density matrix resolved on a fixed $N(t)=n$, and its trace gives the corresponding probability distribution, $\tr\rho(n;t)=P[N(t)=n]$.

The counting-field-resolved density matrix obeys the tilted master equation 
\begin{align}
    \mathcal{L}_\chi \rho_\chi &= \mathcal{L}\rho_\chi + \delta\mathcal{L}_\chi\rho_\chi,\label{eq:tilted-master-equation}\\
    \delta\mathcal{L}_\chi\rho_\chi
    &= \sum_{l=1}^p \left(e^{i\chi\nu_l}-1\right)M_l\rho_\chi M_l^\dagger .\label{eq:tilted-perturbation}
\end{align}
This convention is consistent with the definition of $\rho_\chi$ above: each monitored quantum jump is tracked by a weighted counting field, amounting to the multiplicative factor $e^{i\chi\nu_l}-1$. The cumulant generating function of the integrated current is then
\begin{align}
    C_N(\chi;t) = \ln\tr\rho_\chi(t).
\end{align}
In the long-time limit, the dominant eigenvalue of $\mathcal{L}_\chi$ determines the part of $C_N$ that is linear in time. We therefore define the current cumulant generating function as
\begin{align}
    C_I(\chi) &= \lim_{t\to\infty}\frac{C_N(\chi;t)}{t},\\
    c_n &\equiv \langle\!\langle I^n\rangle\!\rangle
    = (-i\partial_\chi)^n C_I(\chi)\big|_{\chi=0}.\label{eq:fcs-cumulants}
\end{align}
The first cumulant is the mean current, while the second cumulant gives the zero-frequency current noise in the same convention.

For dimensionless discrete counts, it is often useful to work with factorial cumulants instead of ordinary cumulants~\cite{Kambly2011Factorial,Stegmann2015GeneralizedFactorial}. The corresponding generating function is obtained by replacing the exponential counting variable by a factorial counting variable,
\begin{align}
    C_F(z;t) &= C_N(-i\ln(1+z);t),
\end{align}
and the long-time factorial current cumulants are
\begin{align}
    f_n &= \partial_z^n \left[\lim_{t\to\infty}\frac{C_F(z;t)}{t}\right]_{z=0}.
    \label{eq:factorial-cumulants}
\end{align}
Equivalently, the factorial cumulants can be obtained from the ordinary cumulants through the signed Stirling numbers of the first kind,
\begin{align}
    f_m = \sum_{j=1}^m s(m,j)c_j,
    \qquad
    f_1=c_1,\quad f_2=c_2-c_1,\quad f_3=c_3-3c_2+2c_1 .
    \label{eq:factorial-stirling}
\end{align}
In \texttt{QuantumFCS.jl}, this conversion is available through \code{factorial_cumulants}; it can also be requested directly from \code{fcscumulants_recursive} with \code{cumulant_type = "factorial"}. The factorial-cumulant interpretation assumes a dimensionless count variable, so dimensionful heat-current cumulants should be rescaled before applying this conversion.

\begin{figure}[t]
    \centering
    \includegraphics[width=\linewidth]{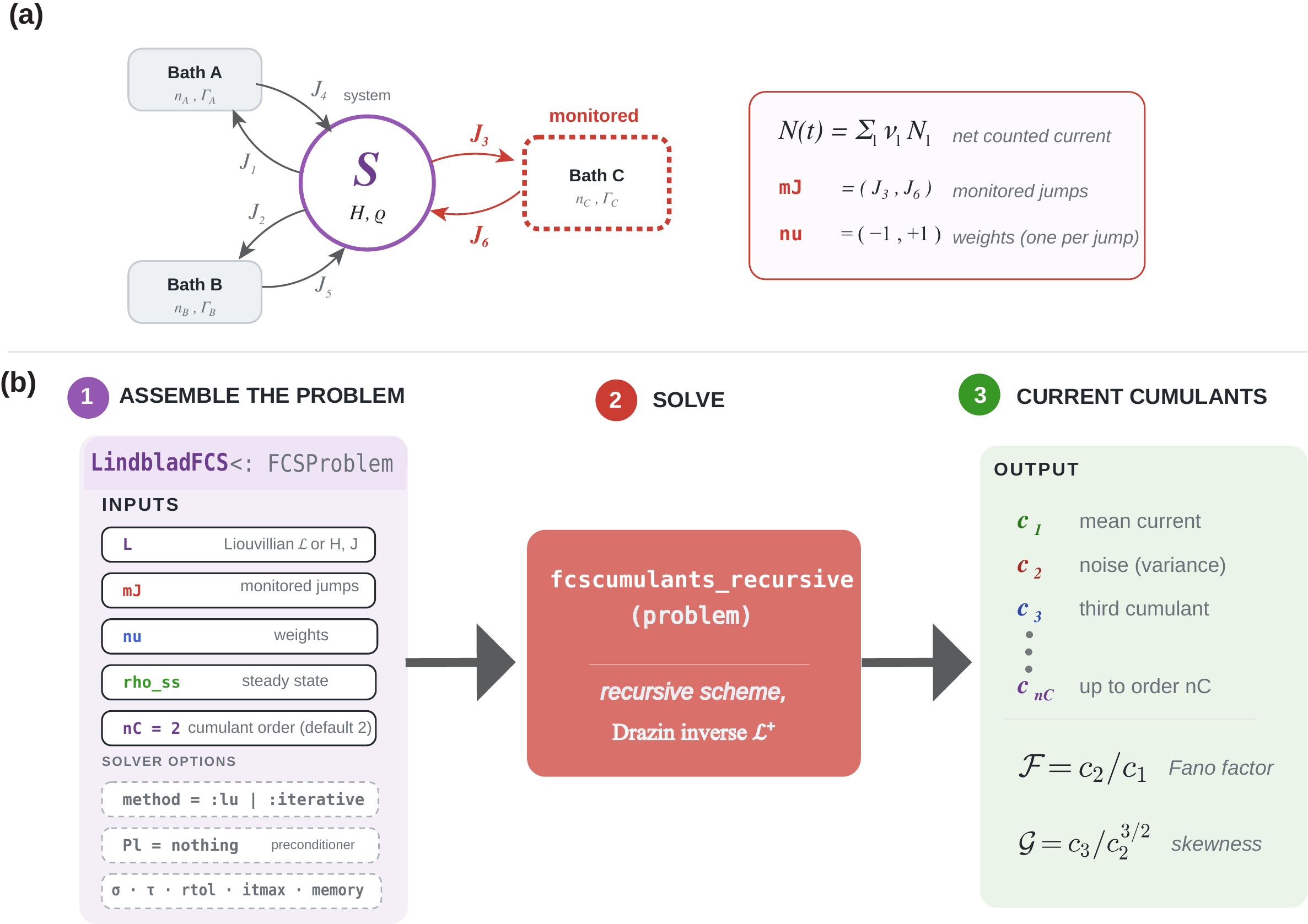}
    \caption{Overview of the FCS workflow implemented in \texttt{QuantumFCS.jl}. \textbf{(a)} An open quantum system is specified by a Hamiltonian $H$ and jump operators $J_k$; a current is defined by selecting monitored jumps $\mathbf{M}$ and weights $\boldsymbol{\nu}$, which determine the integrated current $N(t)=\sum_l\nu_lN_l(t)$. \textbf{(b)} The same data define a \code{LindbladFCS} problem, together with the steady state, cumulant order, and solver options. The function \code{fcscumulants\_recursive} evaluates the recursive Drazin scheme and returns the current cumulants.}
    \label{fig:overview}
\end{figure}

\paragraph{Recursive scheme.}
The long-time formulation gives a direct route to the cumulants without numerical differentiation of the tilted eigenvalue. We use the recursive scheme introduced in Refs.~\cite{Flindt2008,Flindt2010} and reviewed in Ref.~\cite{Landi2024}. In vectorised notation, the cumulants are obtained from
\begin{align}
\langle\!\langle I^n \rangle\!\rangle
&= \sum_{m=1}^n \binom{n}{m}\,
\langle\!\langle 1|\mathcal{L}^{(m)}|\sigma_{n-m}\rangle\!\rangle,\label{eq:cumulant-recursion}\\
|\sigma_n\rangle\!\rangle
&= \mathcal{L}^+ \sum_{m=1}^n \binom{n}{m}
\left(\langle\!\langle I^m\rangle\!\rangle\mathcal{I}-\mathcal{L}^{(m)}\right)
|\sigma_{n-m}\rangle\!\rangle,\label{eq:state-recursion}\\
\mathcal{L}^{(n)}
&= \left(-i\partial_\chi\right)^n \mathcal{L}_\chi\big|_{\chi=0}=\sum_l^p \nu_l^n J^*_l\otimes J_l.\label{eq:tilted-derivatives}
\end{align}
Here, $|A\rangle\!\rangle$ denotes the vectorised form of an operator $A$, $\langle\!\langle 1|$ is the trace functional, $\mathcal{I}$ is the identity super-operator, and $\mathcal{L}^+$ denotes the Drazin inverse of the Liouvillian on the subspace orthogonal to the stationary state. The recursion is initialised with $\sigma_0=\rho_{\mathrm{ss}}$. Equation~\eqref{eq:state-recursion} then determines $\sigma_n$ only from lower-order pseudo-states, so that the cumulants can be generated order by order.

The main numerical bottleneck is the action of the Drazin inverse. Section~\ref{sec:sparse-iterative-backend} explains how this action is
cast as a linear solve in the package interface. The complete workflow, from choosing monitored jumps to computing current cumulants, is summarised in Fig.~\ref{fig:overview}.

\section{Package interface and solver backends}
\label{sec:usage}
This section maps the objects introduced above to the user-facing package interface. We first describe the common problem type and direct function calls, then a minimal transport example that can be read as a complete first workflow, and finally the sparse iterative backend used for larger Liouvillians.

\texttt{QuantumFCS.jl}~\cite{QuantumFCS} is a registered, MIT-licensed Julia package and can be installed from the Julia package manager via
\begin{lstlisting}
using Pkg
Pkg.add("QuantumFCS")
\end{lstlisting}
After installation, the package can be loaded in a Julia script or notebook via
\begin{lstlisting}
using QuantumFCS
\end{lstlisting}

\subsection{Problem type and direct interface}
\label{sec:usage-lindbladfcs}

The package exposes both a direct function interface and a problem-object
interface. As summarised by the input block of Fig.~\ref{fig:overview}(b), the problem object \texttt{LindbladFCS} bundles the Liouvillian data, monitored jumps, weights, steady state, cumulant order, and solver options in a single object. Users may construct it from a Hamiltonian and jump
operators,
\begin{lstlisting}
problem = LindbladFCS(;
    # Model data
    H = H,
    J = J,
    # Current definition and FCS settings
    mJ = mJ,
    rho_ss = rho_ss,
    nu = nu,
    nC = nC,
)
cumulants = fcscumulants_recursive(problem)
\end{lstlisting}
or from an already vectorised Liouvillian,
\begin{lstlisting}
problem = LindbladFCS(;
    # Preassembled Liouvillian data
    L = L,
    # Current definition and FCS settings
    mJ = mJ,
    rho_ss = rho_ss,
    nu = nu,
    nC = nC,
)
cumulants = fcscumulants_recursive(problem)
\end{lstlisting}
The second form is useful when the Liouvillian has already been assembled by an
external package or by problem-specific code. In both cases, \code{mJ} contains
the monitored jump operators, \code{nu} contains the corresponding weights, and
\code{nC} fixes the number of cumulants to compute. The direct function calls
\begin{lstlisting}
fcscumulants_recursive(L, mJ, nC, rho_ss, nu)
fcscumulants_recursive(H, J, mJ, nC, rho_ss, nu)
\end{lstlisting}
are available as well and are equivalent to constructing a \code{LindbladFCS}
problem with a default method based on the LU-factorisation of the Liouvillian.
For dimensionless count observables, factorial cumulants are obtained by passing \code{cumulant_type = "factorial"} to \code{fcscumulants_recursive}, or by applying \code{factorial_cumulants} to a vector of ordinary cumulants.
Backend extensions support \texttt{QuantumOptics.\allowbreak jl}~\cite{QuantumOptics2018}
and \texttt{QuantumToolbox.\allowbreak jl}~\cite{QuantumToolbox2025}. They
extract dense or sparse matrices from those packages' custom types and then
route the calculation to the matching solver. This means users can work directly
with their favourite Julia package for simulating open quantum systems and pass
their objects to the same FCS routine. 

\subsection{Minimal quantum-dot example}
\label{sec:usage-quantum-dot}

We demonstrate the problem-object workflow by studying a single-level quantum
dot coupled to two fermionic reservoirs. A notebook version of this minimal example is available in the accompanying
repository~\cite{QuantumFCSNotebooks}. This system is described by the
Hamiltonian
\begin{align}
    H &= \epsilon_d d^\dagger d
\end{align}
with the energy of the dot $\epsilon_d$ and the mode operator $d$. The dynamics of the system are given by the quantum master equation,
\begin{align}
    \partial_t \rho &= -i \left[ H, \rho\right] + \mathcal{L}_c \rho + \mathcal{L}_h \rho,    \\
    \mathcal{L}_\alpha \rho &= \kappa_\alpha (1 - n_F^\alpha) \mathcal{D}[d] \rho + \kappa_\alpha n_F^\alpha \mathcal{D}[d^\dagger]\rho,
\end{align}
where $n_F^\alpha$ is the Fermi-Dirac occupation of the cold/hot bath.
In the large bias limit electrons only enter the quantum dot from the hot reservoir and leave through the cold reservoir \cite{potts2026quantumthermodynamics}. For the corresponding master-equation FCS of a single-level quantum dot, see Ref.~\cite{PhysRevB.67.085316}; for the thermodynamic formulation of the physical model, see Ref.~\cite{potts2026quantumthermodynamics}, and for an experimental realisation see Ref.~\cite{PhysRevLett.96.076605}.

We compute the first three ordinary cumulants and the corresponding factorial cumulants of the electron count into the cold reservoir.
% For details on the physical model, see Ref.~\cite{potts2026quantumthermodynamics} and for an experimental realisation see Ref.~\cite{PhysRevLett.96.076605}.
Although we consider a minimal single-level model here, the same workflow extends directly to larger Hilbert spaces and multiple monitored transport channels.
To keep the focus of the example on the usage of our package, we use the \texttt{QuantumOptics.jl} package \cite{QuantumOptics2018} to define the model and compute its steady state with its built-in solver. \texttt{QuantumFCS.jl} supports both high-level operator types from \texttt{QuantumOptics.jl} and low-level sparse or dense matrix representations, allowing integration into existing open-quantum-system workflows and flexible integration with raw inputs. The recursive formulation implemented in \code{fcscumulants_recursive} avoids explicit numerical differentiation of the cumulant generating function and enables efficient computation of high-order current cumulants.

We first define the system parameters and construct the operators describing the quantum dot:
\begin{lstlisting}
using QuantumOptics

# Create the basis for the single quantum dot
b = FockBasis(1)

# Define operators
d = destroy(b)
d_dag = create(b)

# System parameters
ϵd = 1.0  # Energy level of the quantum dot
κc = 0.1  # Coupling strength to cold reservoir
κh = 0.5  # Coupling strength to hot reservoir

# Hamiltonian
H = ϵd * d_dag * d

# Jump operators
Jcloss = sqrt(κc) * d

Jhgain = sqrt(κh) * d_dag


J = [Jcloss, Jhgain]

\end{lstlisting}
We now use the built-in solver of the \texttt{QuantumOptics.jl} package to determine the steady state of our system.
\begin{lstlisting}
# Calculate steady state
ρss = steadystate.iterative(H, J)
\end{lstlisting}
% For this example, we monitor electrons entering the cold reservoir.
% To access the cumulants of this current, we select the corresponding jump operator and assign a weight.
Since we monitor the electrons entering the cold reservoir, we select the corresponding jump operators and a positive dimensionless count weight,
\begin{lstlisting}
# Monitored jump operators
mJ = [Jcloss]
# Vector with weights
nu = [+1]
\end{lstlisting}
In the notation of Fig.~\ref{fig:overview}(a), \texttt{mJ} selects the cold-reservoir loss channel as the monitored jump, while $\nu=(+1)$ assigns one positive dimensionless count to each electron leaving the system and entering the cold reservoir.
% Here we compute the first two cumulants of the electron current entering the cold reservoir using \code{fcscumulants_recursive}.

We have now all the necessary inputs to construct a \code{LindbladFCS} problem
and compute the FCS of the cold current:
% The cumulants are then obtained directly from the steady state and the monitored jump operators:
\begin{lstlisting}
# Compute ordinary and factorial current cumulants
problem = LindbladFCS(;
    # Model data from QuantumOptics.jl
    H = H,
    J = J,
    # Cold-reservoir current definition
    mJ = mJ,
    rho_ss = ρss,
    nu = nu,
    nC = 3,
)
ordinary = fcscumulants_recursive(problem)
factorial = fcscumulants_recursive(problem; cumulant_type = "factorial")
println("Ordinary cumulants  = $ordinary")
println("Factorial cumulants = $factorial")
\end{lstlisting}
As indicated by the output block of Fig.~\ref{fig:overview}(b), \texttt{nC = 3} requests the first three cumulants, which are returned in ascending order: $c_1$ is the mean particle current, $c_2$ the zero-frequency current noise, and $c_3$ the third current cumulant. The same problem object can be evaluated with \code{cumulant_type = "factorial"} to return factorial cumulants. The output above is
 \begin{lstlisting}
Ordinary cumulants  = [0.08333333333333334, 0.0601851851851852, 0.03317901234567902]
Factorial cumulants = [0.08333333333333334, -0.02314814814814814, 0.019290123456790098]
\end{lstlisting}
For the large-bias regime considered here, the analytical expressions for the first two cumulants are \cite{potts2026quantumthermodynamics}
\begin{align}
c_1 &= \frac{\kappa_c \kappa_h}{\kappa_c + \kappa_h}, &
c_2 &= \frac{\kappa_h^2 + \kappa_c^2}{(\kappa_c + \kappa_h)^2}\, c_1 .
\end{align}
The numerical results shown above reproduce these analytical predictions with the positive counting convention used here. The nonzero higher factorial cumulants, $f_2=c_2-c_1$ and $f_3=c_3-3c_2+2c_1$, show directly that the electron-transfer statistics is not a Poisson process; for independent Poissonian transfer all factorial cumulants above first order would vanish~\cite{Kambly2011Factorial,Komijani2013HoleTransfer}.

For simplicity, the pseudo-code provided in the examples of Sections~\ref{sec:ddjc} and~\ref{sec:circuit-qhe} does not explicitly show the construction of the model as above. Complete runnable code is found in the accompanying repository~\cite{QuantumFCSNotebooks} for all examples discussed here.
\subsection{Iterative sparse solves}
\label{sec:sparse-iterative-backend}

The default backend applies the Drazin inverse through a sparse/dense LU
factorisation, depending on the input. This is robust for moderate Liouville-space dimensions, but
direct factorisations can become memory-limited for large bosonic Hilbert
spaces. \texttt{QuantumFCS.jl} therefore also provides an iterative Drazin
backend, activated by setting \code{method = :iterative}. This backend uses the
generalised minimal residual method (GMRES)—a subclass of Krylov methods~\cite{Montoison2023}—for the linear systems generated by
the Drazin recursion and preconditions the Krylov iteration with an
incomplete-LU (iLU) factorisation. The extension is activated by loading the
weak dependencies \code{Krylov.jl}~\cite{Montoison2023} and \code{IncompleteLU.jl}; users who only need the
direct backend do not pay this load-time cost.

In this section, a \emph{backend} means the numerical linear-algebra route used
to apply the inverse operations required by the recursion. A
\emph{preconditioner} is an inexpensive approximate inverse that makes the GMRES
iteration converge in fewer steps. A prepared FCS \emph{context} stores the
steady state together with this reusable solver data, so repeated cumulant
calculations do not rebuild the same linear-algebra objects.

The reason this works well is already visible in the vectorised formulation.
The steady state is the right null vector of the Liouvillian,
\begin{align}
    \mathcal{L}|\rho_{\rm ss}\rangle\!\rangle=0,
    \qquad
    \langle\!\langle 1|\rho_{\rm ss}\rangle\!\rangle=1,
    \label{eq:ss-null-vector}
\end{align}
where $\langle\!\langle 1|$ is the trace functional. Numerically, this singular
problem can be replaced by a trace-constrained linear system
\begin{align}
    \left(\mathcal{L}+|w\rangle\!\rangle\langle\!\langle 1|\right)
    |\rho_{\rm ss}\rangle\!\rangle
    =
    |w\rangle\!\rangle ,
    \label{eq:ss-trace-constrained}
\end{align}
where $|w\rangle\!\rangle$ fixes the gauge associated with the null mode.
The steady-state routine \texttt{trace\_constrained\_\allowbreak steadystate}
implements Eq.~\eqref{eq:ss-trace-constrained} for large sparse workflows. It
accepts either a prebuilt Liouvillian or an \code{H}, \code{J} pair and returns
the trace-normalised steady state together with solver diagnostics and reusable
preconditioner data. A lower-level system object stores the trace-constrained
linear system and preconditioner explicitly, which is useful when a parameter
sweep starts a new solve from the previous solution or reuses nearby solver
data.
To connect the same idea to the recursion in Eq.~\eqref{eq:state-recursion}, we define
the source vector at order $n$ as
\begin{align}
    |b_n\rangle\!\rangle
    =
    \sum_{m=1}^n \binom{n}{m}
    \Bigl(
        \langle\!\langle I^m \rangle\!\rangle \mathcal{I}
        -
        \mathcal{L}^{(m)}
    \Bigr)
    |\sigma_{n-m}\rangle\!\rangle ,
    \label{eq:drazin-recursion-source}
\end{align}
where $\mathcal{I}$ is the identity superoperator. The recursion then asks for
the Drazin action
$|\sigma_n\rangle\!\rangle=\mathcal{L}^+|b_n\rangle\!\rangle$. Following the
linear-system formulation reviewed in Ref.~\cite{Landi2024}, this action can be
obtained by solving for $|y_n\rangle\!\rangle=|\sigma_n\rangle\!\rangle$ in
\begin{align}
    \left(\mathcal{L}
    +
    |\rho_{\rm ss}\rangle\!\rangle\langle\!\langle 1|\right)
    |y_n\rangle\!\rangle
    =
    Q|b_n\rangle\!\rangle,
    \qquad
    Q=\mathcal{I}-|\rho_{\rm ss}\rangle\!\rangle\langle\!\langle 1|,
    \label{eq:drazin-gauge-fixed}
\end{align}
with the trace gauge $\langle\!\langle 1|y_n\rangle\!\rangle=0$. The projector
$Q$ removes the stationary component of the source, while the trace gauge fixes
the otherwise non-unique inverse on the subspace orthogonal to the steady state.
Thus the steady-state solve and the Drazin solves are not unrelated tasks: they
are linear systems built from the same sparse Liouvillian, differing only by the
trace/gauge terms that remove the null mode.

For large sparse problems, the steady state is often found with GMRES and an
iLU preconditioner for Eq.~\eqref{eq:ss-trace-constrained}. The
\code{trace\_constrained\_steadystate} call returns an object containing the trace-normalised
steady state \code{ss.rho\_ss}, the sparse Liouvillian \code{ss.L}, scalar
diagnostics \code{ss.stats}, and the shifted-iLU preconditioner \code{ss.Pl}. If this
preconditioner approximates the shifted Liouvillian well, it is also a natural
preconditioner for the Drazin solve in Eq.~\eqref{eq:drazin-gauge-fixed}. Passing it through
\code{prepare\_fcs\_context} or the \code{Pl} field skips a second iLU
construction,
\begin{lstlisting}
using Krylov, IncompleteLU

# Solve the trace-constrained steady-state equation with GMRES + iLU
ss = trace_constrained_steadystate(L; method = :iterative)
# Reuse the steady state and preconditioner for FCS cumulants
ctx = prepare_fcs_context(ss; method = :iterative)
cumulants = fcscumulants_recursive(ctx; mJ = mJ, nu = nu, nC = nC)
\end{lstlisting}
The bridge \code{prepare\_fcs\_context(ss; method = :iterative)} forwards
\code{ss.Pl} to the iterative Drazin backend, so the FCS calculation does not
rebuild the iLU. The supplied preconditioner is applied on the right, so GMRES
is still stopped using the true residual of the Drazin linear system.
Section~\ref{sec:ddjc} uses this pattern in a parameter sweep where the
Liouvillian dimension exceeds $10^6$ for some parameter regions.

\section{Driven-dissipative Jaynes-Cummings model}
\label{sec:ddjc}

We next consider the driven-dissipative Jaynes-Cummings model \cite{Jaynes:1963zz} in the regime where the
photon blockade \cite{PhysRevLett.79.1467,Birnbaum2005} breaks down, as analysed by
Carmichael~\cite{Carmichael2015BreakdownPhotonBlockade}. This example is useful
for testing full-counting-statistics methods because the same ingredients that
make the physics interesting--strong coupling, strong drive, and slow
switching between dim and bright states--also make the numerical problem large
and ill-conditioned.

\subsection{Model and photon statistics}
We work in the rotating frame of the coherent drive and set $\hbar=1$. The
Hamiltonian is
\begin{align}
    H_{\rm JC}
    =
    -\Delta(a^\dagger a+\sigma_+\sigma_-)
    +g(a^\dagger\sigma_-+a\sigma_+)
    -\mathcal{E}(a+a^\dagger),
    \label{eq:ddjc-hamiltonian}
\end{align}
where $a$ annihilates a cavity photon, $\sigma_-$ lowers the two-level system,
$g$ is the Jaynes-Cummings coupling, $\mathcal{E}$ is the coherent-drive amplitude, and
$\Delta=\omega_D-\omega_0$ is the drive detuning from the resonant cavity and
two-level transition frequency. The sign of the last term fixes the phase of
the drive and does not affect the photon-number observables considered below.

The open-system dynamics is generated by cavity loss only, with no spontaneous
emission from the two-level system,
\begin{align}
    \dv{\rho}{t}
    =
    -i[H_{\rm JC},\rho]
    +\kappa\mathcal{D}[a]\rho,
    \label{eq:ddjc-master-equation}
\end{align}
where $\mathcal{D}$ is the Lindblad dissipator of Eq.~\eqref{eq:LME}. Thus
$\kappa$ is the photon-number decay rate used in the numerical implementation.
The counted process is the positive cavity-emission current, associated with the
monitored jump operator
\begin{align}
    J=\sqrt{\kappa}\,a,
    \qquad
    \nu=1 .
    \label{eq:ddjc-counted-jump}
\end{align}
Consequently, the first long-time cumulant is
\begin{align}
    c_1=\kappa\langle a^\dagger a\rangle_{\rm ss},
    \label{eq:ddjc-mean-count}
\end{align}
and we characterise the output statistics using the Fano factor
\begin{align}
    \mathcal{F}=\frac{c_2}{c_1}
    \label{eq:ddjc-fano}
\end{align}
together with the normalised third cumulant 
\begin{equation}
     \mathcal{G}=\frac{c_3}{c_2^{3/2}}.
\end{equation}
While Ref.~\cite{Carmichael2015BreakdownPhotonBlockade} analyses steady-state
properties, the FCS calculations below probe the photon statistics across the
crossover between blockaded and bright emission.

\subsection{Blockade-breakdown crossover regime}
The relevant regime is specified by three dimensionless parameters:
\begin{align}
    \frac{2\mathcal{E}}{g},
    \qquad
    \frac{\Delta}{\kappa},
    \qquad
    \frac{g}{\kappa} .
    \label{eq:ddjc-dimensionless-parameters}
\end{align}
Physically, these tune the cavity between two regimes. At weak drive the strong
Jaynes-Cummings non-linearity prevents more than a few photons from accumulating---the
\emph{photon blockade}---and the cavity sits in a nearly empty, \textit{dim} state; at strong
drive the blockade breaks down and the cavity fills to a high-photon, \textit{bright} state.
The physics of interest is the blockade-breakdown crossover between these dim and bright
states, and how sharp it is. Here $2\mathcal{E}/g$ measures the distance from the resonant
breakdown point, $\Delta/\kappa$ selects resonant or detuned sectors of the dressed-state
ladder, and $g/\kappa$ sets the effective photon-number scale.

The blockade originates in the anharmonic Jaynes-Cummings ladder. Under a resonant drive
its quasienergy doublets are
\begin{align}
    E_{n,\pm}(\Delta = 0)
    =
    \pm\sqrt{n}g\left[1-(2\mathcal{E}/g)^2\right]^{3/4},
    \label{eq:ddjc-quasienergy-collapse}
\end{align}
which at zero drive reduce to the bare splitting $\pm\sqrt{n}g$. This $\sqrt{n}$ spacing
makes successive transitions unequal, so a drive resonant with one rung is detuned from
the next---the blockade---by an amount that scales as $g/\sqrt{n}$ and therefore weakens
at high photon number~\cite{Carmichael2015BreakdownPhotonBlockade}. Under drive the quasi-energies coalesce exactly at $2\mathcal{E}/g=1$: there the ladder is
equally spaced and the blockade breaks down. This fixes the organising centre of the
transition from the Hamiltonian alone, but not how sharp the crossover is.

Its sharpness is set by how many photons the bright state carries. A quick estimate
follows from the cavity mode alone: driven at $\mathcal{E}$ and damped at $\kappa$ it
would settle at $\langle a^\dagger a\rangle=(2\mathcal{E}/\kappa)^2$, i.e. at criticality
\begin{align}
    n_{\rm scale}=\left(\frac{g}{\kappa}\right)^2.
\end{align}
The empty cavity has no transition of its own, so this is only an order-of-magnitude
occupation scale; it nonetheless plays the role of an effective system size. Increasing
$g/\kappa$ raises $n_{\rm scale}$ and sharpens the crossover, which becomes a genuine
dissipative phase transition as $g/\kappa\to\infty$---the special strong-coupling
``thermodynamic limit'' of Ref.~\cite{Carmichael2015BreakdownPhotonBlockade}.

The nature of the crossover depends on the detuning. On resonance it is continuous. Away
from resonance the upper and lower ladders separate, with a local transition detuning
$\simeq-\Delta\pm g/(2\sqrt{n})$ (a large-$n$ mean-field estimate; see
App.~\ref{app:ddjc-cutoff}). Positive detuning then selects the upper-ladder channel: the dim and
bright states \emph{coexist} over a finite drive window, a first-order-like bistability
along a line in the $(2\mathcal{E}/g,\Delta/\kappa)$ plane that the cuts
$\Delta/\kappa=0.55$ and $0.70$ used in the plots of the following section cross. Below we show that full-counting statistics, and in
particular the skewness $\mathcal{G}$, characterises this dim--bright crossover directly
from the emitted light.

\subsection{Full-counting statistics in the blockade-breakdown regime}

Numerically, the regime described above is substantially more costly than the
minimal quantum-dot example. We use fixed $g/\kappa=14$ and sweep the drive
ratio $2\mathcal{E}/g$ from $0.05$ to $1.45$ for
$\Delta/\kappa\in\{0,0.55,0.70\}$. At the resonant critical drive
$2\mathcal{E}/g=1$, the effective system size is
$n_{\rm scale}=196$ (defined above). A Fock cutoff $N$ gives Hilbert-space
dimension $2(N+1)$ and vectorised Liouville-space dimension $[2(N+1)]^2$; at
the largest cutoff used here, $N=500$, the sparse non-Hermitian Liouvillian has
dimension $1004^2$. Dense methods and repeated sparse direct factorisations are
therefore impractical on a tabletop machine; here, we use a MacBook Pro M1 with
16 GB RAM, while Ref.~\cite{Carmichael2015BreakdownPhotonBlockade} uses a
dedicated cluster\footnote{Note that here we downscale the problem from $g/\kappa = 28$ (Ref.~\cite{Carmichael2015BreakdownPhotonBlockade}) to $g/\kappa =14$; however, FCS is in principle much more demanding than simply steady-state solving. Reference~\cite{Carmichael2015BreakdownPhotonBlockade} suggests that at least $N=900 \approx 28^2$ were needed for steady-state solutions, since at $N=900$ the mean occupation approaches the cutoff, $\langle a^\dagger a\rangle_\text{ss}\approx N$ (occupation fraction $\approx 1$), which is not cutoff safe for the parameter regions exhibiting crossover. A direct comparison is not possible since the cutoff used and underlying method settings are not provided.}.

We therefore use the sparse iterative strategy described in
Sec.~\ref{sec:sparse-iterative-backend} (trace-constrained GMRES with a
shifted-iLU preconditioner, reused as the Drazin right-preconditioner). The Fock
cutoffs are selected dynamically, $N=150,\ldots,500$, from a semi-classical
approximation and validated by the diagnostics in
Fig.~\ref{fig:ddjc-fcs-scans}(g--i). The cutoff rule is detailed in
Appendix~\ref{app:ddjc-cutoff}, the preconditioned sparse solves in
Appendix~\ref{app:ddjc-preconditioner}, and the validation checks in
Appendix~\ref{app:ddjc-validation}.

\begin{figure}[t]
    \centering
   \includegraphics[width = \linewidth]{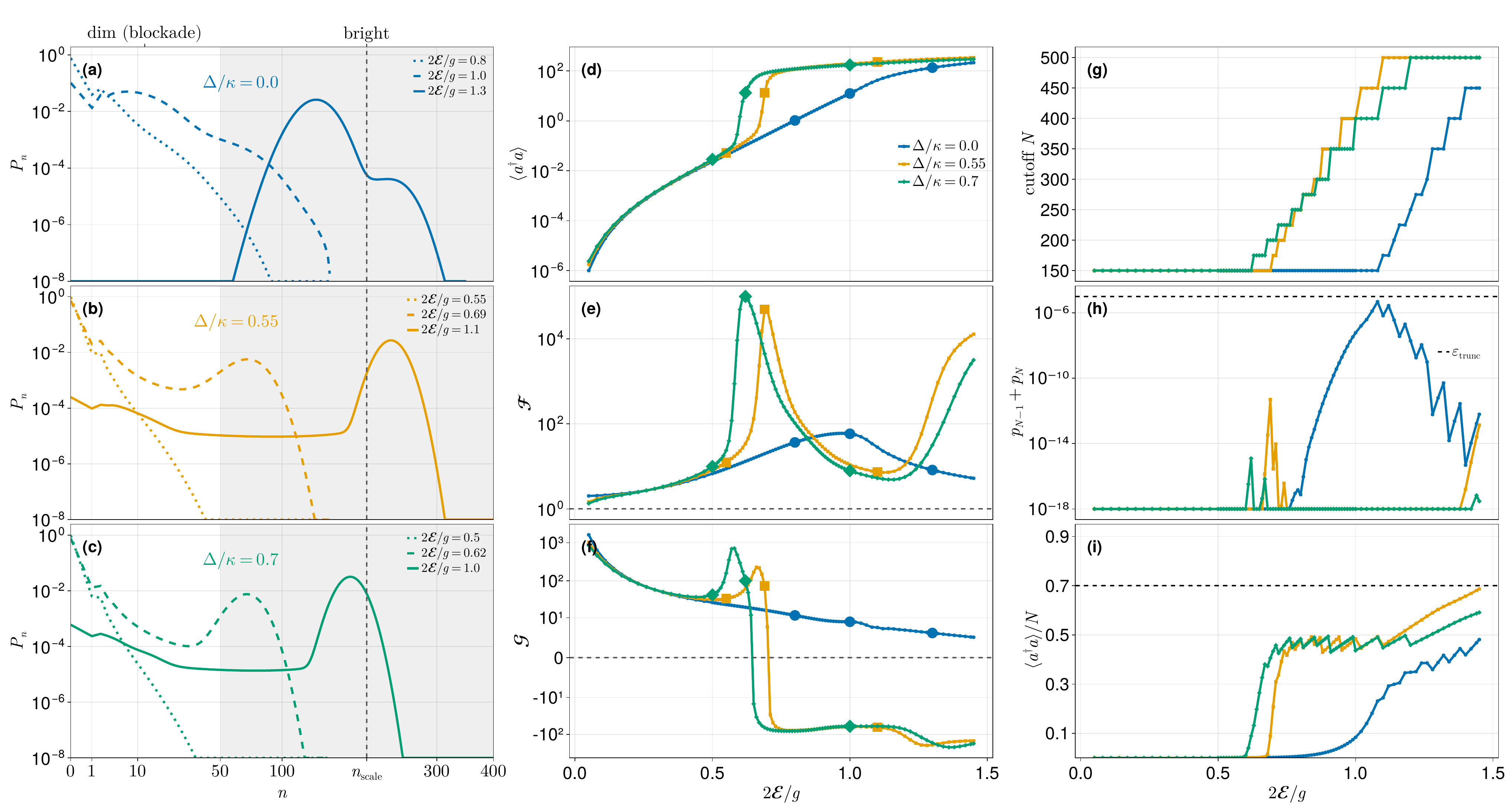}
    \caption{
    Full-counting statistics and cavity Fock-state distributions of the
    driven-dissipative Jaynes-Cummings model at fixed $g/\kappa=14$ and
    $\Delta/\kappa\in\{0,0.55,0.70\}$ (blue, orange, green), swept as a function
    of $2\mathcal{E}/g$.
    \textbf{(a--c)} Steady-state cavity Fock-state distribution $P_n$ for each
    detuning, at three drives spanning that cut's blockade-breakdown transition
    (below, at, and after: dotted, dashed, solid; values in the legends). $P_n$
    is on a logarithmic scale and the photon number $n$ on a square-root scale
    to resolve the dim peak near $n=0$. The vertical dashed line and the
    bottom-axis $n_{\rm scale}$ tick mark the semi-classical scale
    $n_{\rm scale}=(g/\kappa)^2=196$; the shaded band and the top-axis labels
    indicate the dim (blockade) and bright regions.
    \textbf{(d--f)} The mean photon number
    $c_1/\kappa=\langle a^\dagger a\rangle_{\rm ss}$, the Fano factor
    $\mathcal{F}=c_2/c_1$, and the normalised third cumulant
    $\mathcal{G}=c_3/c_2^{3/2}$; enlarged markers flag the drives whose $P_n$
    are shown in (a--c).
    \textbf{(g--i)} Numerical-reliability diagnostics for the same runs: the
    dynamically selected Fock cutoff $N$, the boundary population $p_{N-1}+p_N$
    with the truncation threshold $\epsilon_{\rm trunc}=10^{-5}$, and the
    occupation fraction $\langle a^\dagger a\rangle_{\rm ss}/N$.
    }
    \label{fig:ddjc-fcs-scans}
\end{figure}

The corresponding package call for these simulations has the following form:
% using QuantumToolbox

% # QuantumToolbox.jl assembles the sparse Liouvillian
% L = QuantumToolbox.liouvillian(H, [cavity_loss])
\begin{lstlisting}
using Krylov, IncompleteLU
# QuantumFCS.jl solves the steady state with GMRES + iLU
ss = trace_constrained_steadystate(L; method = :iterative)

# Reuse the steady-state preconditioner for the FCS recursion
problem = LindbladFCS(;
    L = ss.L,
    mJ = [cavity_loss],
    rho_ss = ss.rho_ss,
    nu = [1.0],
    nC = 3,
    method = :iterative,
    Pl = ss.Pl,
    rtol = 1e-8,
    itmax = 300,
    memory = 60,
)
c1, c2, c3 = fcscumulants_recursive(problem)
\end{lstlisting}
The drive sweep uses application-level continuation,
meaning that neighbouring parameter points reuse previous solver information
when this remains numerically reliable.
It additionally applies cutoff
selection and validation policies. 
 %\texttt{QuantumFCS.jl} is designed so that continuation  but the single-point trace-constrained
% solve and preconditioner reuse are part of \texttt{QuantumFCS.jl}.
The trace-constrained linear solve from \texttt{QuantumFCS.jl} is precisely designed to allow such continuation and monitoring of solver reliability, useful for numerically demanding applications.
Note that here we did not set up a context; this feature is particularly useful in Section~\ref{sec:circuit-qhe}, where we will need to compute FCS of multiple currents.

Figure~\ref{fig:ddjc-fcs-scans}(a--c) shows the steady-state cavity Fock distributions
$P_n$ directly. On resonance [Fig.~\ref{fig:ddjc-fcs-scans}(a)] a single peak moves
continuously from the dim region ($n\ll n_{\rm scale}$) to the bright region as
$2\mathcal{E}/g$ crosses one; for the detuned cuts
[Fig.~\ref{fig:ddjc-fcs-scans}(b,c)] two peaks, one dim and one bright, are
simultaneously populated---the hallmark of the coexistence region introduced above. This
distribution already contains the physics of the crossover, but reconstructing it
experimentally would require tomography of a cavity state spanning hundreds of Fock
levels. The counting statistics of the emitted light
[Fig.~\ref{fig:ddjc-fcs-scans}(d--f)], obtained directly from photocounts, give an
experimentally accessible fingerprint of the same dim--bright structure, as we now show.

The resonant and detuned cuts probe different aspects of the blockade breakdown. At
$\Delta=0$, the mean photon number in Fig.~\ref{fig:ddjc-fcs-scans}(d) remains small
below the critical value $2\mathcal{E}/g=1$ and then crosses smoothly into the bright
region, tracking the single shifting peak of Fig.~\ref{fig:ddjc-fcs-scans}(a). The Fano
factor in Fig.~\ref{fig:ddjc-fcs-scans}(e) reaches a moderate super-Poissonian maximum
near the crossover and then decreases as the bright state becomes better developed,
although it remains above the Poisson value.

The detuned cuts show a much sharper response. For $\Delta/\kappa=0.55$ and
$0.70$, the upper-ladder channel of
Eq.~\eqref{eq:ddjc-large-n-detuning} is selected, and the mean photon number in Fig.~\ref{fig:ddjc-fcs-scans}(d) then jumps by
orders of magnitude over a narrow drive window, around
$2\mathcal{E}/g\simeq0.69$ for $\Delta/\kappa=0.55$ and
$2\mathcal{E}/g\simeq0.62$ for $\Delta/\kappa=0.70$. Exactly at these jumps,
the Fano factor in Fig.~\ref{fig:ddjc-fcs-scans}(e) reaches values of order $10^5$, far above coherent-light and
ordinary thermal-bunching scales. This reflects telegraph-like switching in the emission from the cavity state, which at the selected detunings is bimodal in the Fock distribution [Fig.~\ref{fig:ddjc-fcs-scans}(b,c)], to which the Fano factor is highly sensitive.
The Fano factor therefore does more than
quantify excess noise: it identifies the slow switching between dim and bright states,
which is only weakly visible in the mean current alone.

The skewness, $\mathcal{G}$, in Fig.~\ref{fig:ddjc-fcs-scans}~(f) resolves the most interesting aspect of the quantum state of light hosted inside the cavity: which phase dominates—dim or bright? For the resonant curve, the continuous transition between dim and bright phases never leads to bimodality of the cavity state; sweeping $\mathcal{G}>0$ indicates that photon emission is dominated by events below the average current, and tends to a symmetric regime past $2\mathcal{E}/g=1$. For the detuned cuts, the cavity state is bimodal [Fig.~\ref{fig:ddjc-fcs-scans}(b,c)]. Regions with $\mathcal{G}>0$ indicate the dim phase dominates, consistent with a mostly blockaded
output punctuated by rare bright bursts. $\mathcal{G}=0$ identifies the point at which dim and bright phases contribute equally to the photo-counting.
Just beyond the dip,
$\mathcal{G}<0$: the bright output has become dominant and is
now interrupted by comparatively rare dim intervals. The sign of
$\mathcal{G}$ thus distinguishes the two sides of the coexisting
blockade-breakdown regime. Figures~\ref{fig:ddjc-fcs-scans}(g--i) show that this
interpretation is not an artefact of visible cutoff leakage or failed solver convergence.

On the one hand, our results show that \texttt{QuantumFCS.jl} allows efficient simulation of the driven-dissipative Jaynes-Cummings model in a physically non-trivial and numerically challenging regime. On the other hand, they also showcase that FCS gives access to quantities which characterise central aspects of the cavity state, which would in principle require quantum state tomography to probe. 
% Altogether~$\mathcal{F}$ and $\mathcal{G}$ allow picturing the dynamic process of photon emission for the detuned cuts $\Delta/\kappa>0$ around $2\mathcal{E}/g = 1$. Strong super-Poissonian behaviour is captured by $\mathcal{F}\gg 1$ for both $2\mathcal{E}/g<1$, and $2\mathcal{E}/g>1$—whenever photons are emitted, they come in bursts containing a large amounts of them. But for $2\mathcal{E}/g<1$, $\mathcal{G}>0$ indicates these bursts are rare (dim, blockaded branch) while for $2\mathcal{E}/g>1$, $\mathcal{G}<0$ indicates bursts are frequent (bright branch).

\section{Quantum heat engine based on Cooper-pair tunnelling}
\label{sec:circuit-qhe}
We now turn to a quantum heat engine based on photon-assisted Cooper-pair tunnelling in a circuit-QED architecture~\cite{HoferSouquetClerk2016}. The model is useful for our purposes because it combines several features that are common in quantum-thermodynamic applications of FCS. It involves heat currents rather than particle currents, bosonic Hilbert spaces that require a truncation, and a non-linear interaction whose current fluctuations are not available from simple Gaussian formulae. It therefore illustrates how \texttt{QuantumFCS.jl} can be used to study non-Gaussian quantum heat engines beyond analytically solvable transport examples.
The section is organised around three package-relevant tasks: defining heat currents through monitored thermal jumps and energy weights, reusing the same Liouvillian context for different currents, and checking truncation and rotating-wave consistency in a non-linear bosonic model.

The system consists of two bosonic modes, labelled hot and cold, with frequencies $\Omega_h$ and $\Omega_c$. Their annihilation operators are denoted by $a_h$ and $a_c$. In the laboratory frame the Hamiltonian reads
\begin{align}
    H(t) &= \Omega_h a_h^\dagger a_h+\Omega_c a_c^\dagger a_c-E_J\cos \varphi(t),
    \label{eq:qhe-lab-hamiltonian}\\
    \varphi(t) &= 2\left(eVt+\Phi_h+\Phi_c\right),
\end{align}
where $E_J$ is the Josephson energy, $V$ is the voltage bias, and
\begin{align}
    \Phi_\alpha = \lambda_\alpha(a_\alpha^\dagger+a_\alpha), \qquad \alpha\in\{h,c\},
\end{align}
with $\lambda_\alpha$ the zero-point phase fluctuations. Dissipation is modelled by local thermal baths with occupations $\bar n_\alpha$ and rates $\kappa_\alpha$; the corresponding jump operators are introduced below in the FCS construction.

\subsection{Rotating-wave model}
Many photon-assisted processes appear in the Josephson term of
Eq.~\eqref{eq:qhe-lab-hamiltonian}. A time-independent effective model is
obtained by selecting processes that satisfy the resonance condition
\begin{align}
    2eV = k\Omega_h-l\Omega_c,
    \qquad k,l\in \mathbb{Z}.
    \label{eq:qhe-resonance}
\end{align}
The corresponding rotating-wave Hamiltonian $H_{k,l}$ is in the rotating frame with respect to the drive,
\begin{align}
    H_{k,l}
    =
    -\frac{E_J}{2}\left[
    i^{k+l}(a_c^\dagger)^l A_c(l)A_h(k)a_h^k+\mathrm{H.c.}
    \right],
    \label{eq:qhe-general-rwa}
\end{align}
where we introduced the Laguerre operators
\begin{align}
    A_\alpha(k)
    &=
    (2\lambda_\alpha)^k e^{-2\lambda_\alpha^2}
    \sum_{n=0}^{\infty}
    \frac{n!}{(n+k)!}
    L_n^{(k)}(4\lambda_\alpha^2)\ket{n}\!\bra{n},
    \label{eq:qhe-laguerre-operator}
\end{align}
which are diagonal in the Fock basis, with $L_n^{(k)}$ the generalised Laguerre polynomial (details in App.~\ref{app:qhe-rwa}); in the
numerical implementation this diagonal operator is built directly in the truncated Fock basis
[Eq.~\eqref{eq:qhe-non-linear-hamiltonian}].

% The solver route is that of Sec.~\ref{sec:sparse-iterative-backend}; only the model
% construction and the two-current context reuse are specific here, and are omitted. Below, we demonstrate how a predefined context \code{ctx} is reused to compute multiple thermodynamic current FCS.
% using QuantumToolbox

% # QuantumToolbox.jl operators define the truncated bosonic model
% ah = tensor(destroy(Nh), qeye(Nc))
% ac = tensor(qeye(Nh), destroy(Nc))

% # Pseudocode: build the RWA Hamiltonian and thermal jumps
% H, J = nonlinear_heat_engine_model(ah, ac; lambda_h, lambda_c, g)
% # QuantumToolbox.jl assembles the Liouvillian matrix
% L = QuantumToolbox.liouvillian(H, J)
% \begin{lstlisting}
% using Krylov, IncompleteLU

% # QuantumFCS.jl solves the trace-constrained steady state
% ss = trace_constrained_steadystate(L; method = :iterative)

% # Reuse the steady-state data for multiple monitored currents
% ctx = prepare_fcs_context(ss; method = :iterative)
% \end{lstlisting}
% Here the pseudo-function \code{nonlinear\_heat\_engine\_model} stands for the RWA construction
% (App.~\ref{app:qhe-rwa}), specialised to the $(1,2)$ process of
% Eq.~\eqref{eq:qhe-non-linear-hamiltonian}. The
% prepared context is then reused for both heat currents, carrying both the
% steady state and the steady-state preconditioner.

The linear process $k=l=1$ provides a useful reference case: in the low-impedance limit the model reduces to a Gaussian heat engine with closed-form expressions for the mean current and scaled variance~\cite{Kerremans2022,Janovitch2023}. We use this limit as a consistency check for the numerical FCS routine before studying genuinely non-linear processes, and in Section~\ref{sec:benchmarking} for benchmarking against analytical results. We now study a non-Gaussian process.

\subsection{Non-linear heat-engine process}
The application considered here selects the process $k=1,l=2$, corresponding to the conversion of one hot photon into two cold photons, or the reverse process. This process is non-Gaussian, and does not lead to a closed linear system of equations. The voltage is fixed to
\begin{align}
    2eV = \Omega_h-2\Omega_c,
    \label{eq:qhe-non-linear-voltage}
\end{align}
and the effective Hamiltonian becomes
\begin{align}
    H_{\rm nl}
    =
    \frac{E_J}{2}\left[
    i(a_c^\dagger)^2 A_c(2)A_h(1)a_h+\mathrm{H.c.}
    \right],
    \label{eq:qhe-non-linear-hamiltonian}
\end{align}
The conditions under which the above RWA Hamiltonian is valid are discussed in Appendix~\ref{app:qhe-rwa}, following Ref.~\cite{HoferSouquetClerk2016}.
The Laguerre operators $A_h(1)$ and $A_c(2)$ encode blockade-related matrix elements inherited from the Josephson interaction. They can strongly modify the access to different Fock states and thereby affect both the average heat current and its fluctuations.
In the numerical scans below we parametrise the Josephson strength by the effective scale 
\begin{align}
    g=\frac{E_J}{2}(2\lambda_h)^k(2\lambda_c)^l= 4E_J\lambda_h\lambda_c^2.\label{eq:qhe-non-linear_coupling}
\end{align}
% while retaining the full Laguerre operators in Eq.~\eqref{eq:qhe-non-linear-hamiltonian}.

The local thermal jump operators are ordered as
\begin{align}
    \mathbf{J}
    =
    \bigl(
    \sqrt{(\bar n_h+1)\kappa_h}\,a_h,\,
    \sqrt{(\bar n_c+1)\kappa_c}\,a_c,\,
    \sqrt{\bar n_h\kappa_h}\,a_h^\dagger,\,
    \sqrt{\bar n_c\kappa_c}\,a_c^\dagger
    \bigr).
    \label{eq:qhe-jump-ordering}
\end{align}
Following the monitored-jump convention of Sec.~\ref{sec:theory}, the hot heat current is obtained from $(J_1,J_3)$ with weights $(-\Omega_h,\Omega_h)$, while the cold heat current is obtained from $(J_2,J_4)$ with weights $(-\Omega_c,\Omega_c)$.

The solver route is that of Sec.~\ref{sec:sparse-iterative-backend}; only the model
construction and the two-current context reuse are specific here, and their construction is omitted. Below, we demonstrate how a predefined context \code{ctx} is used to compute the FCS of both relevant thermodynamic currents, 
\begin{lstlisting}
# Hot heat current: loss/gain jumps weighted by energy
hot = fcscumulants_recursive(ctx;
    mJ = [J[1], J[3]], nu = [-Ωh, Ωh], nC = 2)
# Cold heat current: loss/gain jumps weighted by energy
cold = fcscumulants_recursive(ctx;
    mJ = [J[2], J[4]], nu = [-Ωc, Ωc], nC = 2)
\end{lstlisting}
Here, \code{hot[1]} and \code{cold[1]} are $\langle J_h\rangle$ and
$\langle J_c\rangle$, while the second entries are the corresponding
zero-frequency heat-current noises. The average output power shown in the plots is
\begin{align}
    \langle P\rangle = -\langle J_h\rangle-\langle J_c\rangle .
    \label{eq:qhe-power}
\end{align}

The entropy production rate can be written directly in terms of the thermodynamic affinity per elementary $(1,2)$ conversion event,
\begin{align}
    \sigma
    =
    2\frac{\langle J_h\rangle}{\Omega_h}\mathcal{A}
    =
    -\frac{\langle J_c\rangle}{\Omega_c}\mathcal{A},
    \label{eq:qhe-entropy-affinity}
\end{align}
where
\begin{align}
    \mathcal{A}
    =
    \ln(1+\bar n_c^{-1})-\frac{1}{2}\ln(1+\bar n_h^{-1}) .
    \label{eq:qhe-affinity}
\end{align}
This form uses the stationary tight-coupling relation $2\Omega_c\langle J_h\rangle=-\Omega_h\langle J_c\rangle$ derived in Appendix~\ref{app:qhe-thermo}. In the heat engine regime, $\langle J_h\rangle>0$ and $\langle J_c\rangle<0$.
The Carnot, or reversible equilibrium, point corresponds to $\mathcal{A}=0$.
For fixed $\bar n_h=0.5$ this gives $\bar n_c^{\rm eq}=1/(\sqrt{1+\bar n_h^{-1}}-1)\simeq1.366$.

\subsection{Heat-current fluctuations}
\begin{figure}[t]
    \centering
    \includegraphics[width=\linewidth]{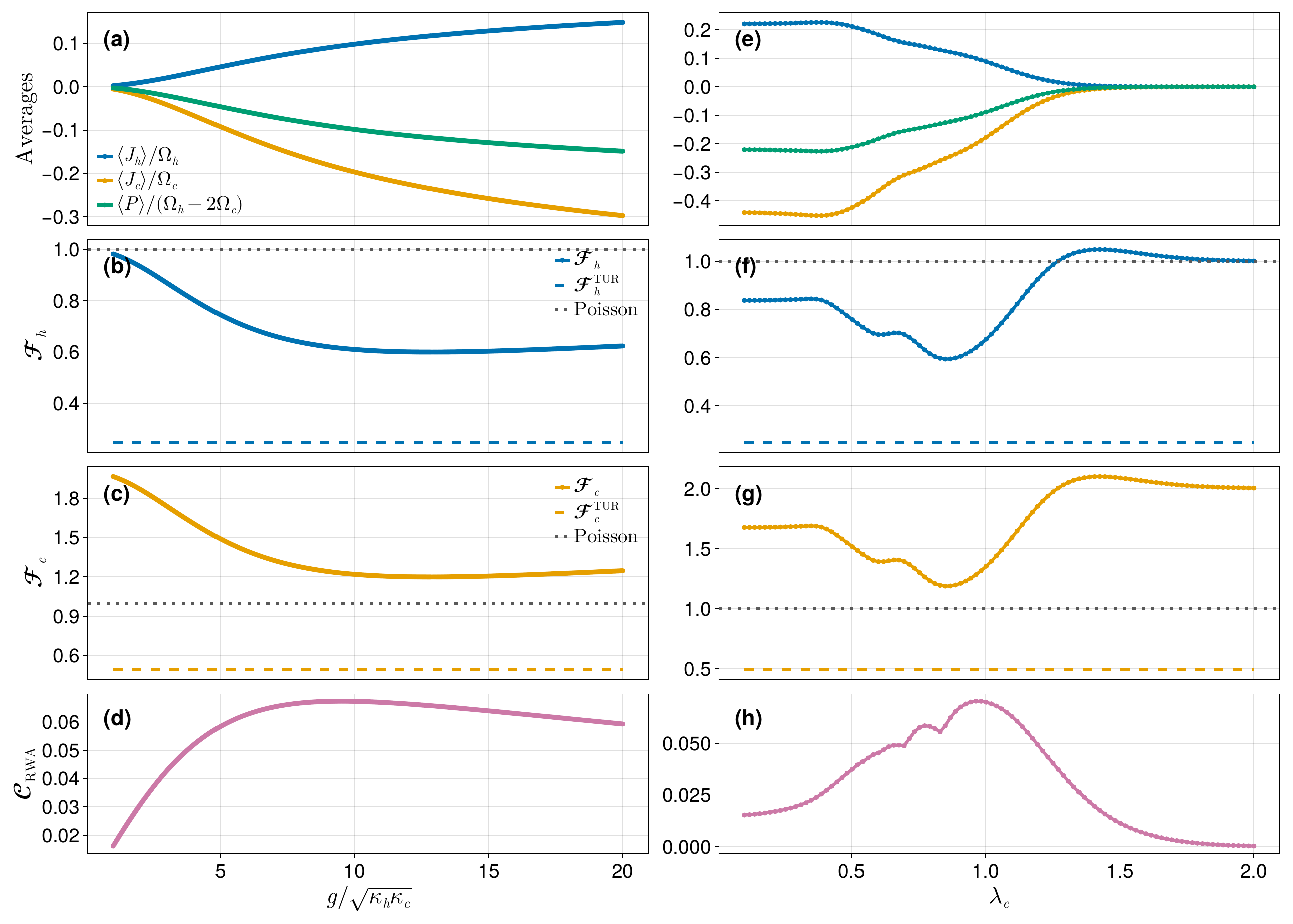}
    \captionof{figure}{
    Antibunching regime of the non-linear heat engine. The left column shows the $g$ sweep at fixed $\lambda_c=0.89$, while the right column shows the $\lambda_c$ sweep at the strongest viable hot-antibunching point of the $g$ scan, $g\simeq7.21$. Panels \textbf{(a)} and \textbf{(e)} show the scaled heat currents and output power, \textbf{(b)} and \textbf{(f)} the hot Fano factor with its TUR threshold and the Poisson reference, \textbf{(c)} and \textbf{(g)} the corresponding cold Fano factor, and \textbf{(d)} and \textbf{(h)} the RWA-link coherence $\mathcal{C}_{\rm RWA}$. The sweep is centred around the blockade-like regime selected by the Laguerre matrix elements.
    }
    \label{fig:qhe-antibunching}
\end{figure}

The dimensionless Fano factors for the hot and cold heat currents are
\begin{align}
    \mathcal{F}_h
    =
    \frac{\langle\!\langle J_h^2\rangle\!\rangle}{\Omega_h|\langle J_h\rangle|},
    \qquad
    \mathcal{F}_c
    =
    \frac{\langle\!\langle J_c^2\rangle\!\rangle}{\Omega_c|\langle J_c\rangle|}.
    \label{eq:qhe-fano}
\end{align}
Values $\mathcal{F}_\alpha<1$ indicate sub-Poissonian heat-transfer statistics in channel $\alpha$. The same FCS data determine the thermodynamic uncertainty products
\begin{align}
    \mathcal{Q}_h
    =
    \frac{\langle\!\langle J_h^2\rangle\!\rangle}{\langle J_h\rangle^2}\sigma
    =
    2\mathcal{A}\mathcal{F}_h,
    \qquad
    \mathcal{Q}_c
    =
    \frac{\langle\!\langle J_c^2\rangle\!\rangle}{\langle J_c\rangle^2}\sigma
    =
    \mathcal{A}\mathcal{F}_c .
    \label{eq:qhe-tur-product}
\end{align}
Classical Markovian dynamics obeys the thermodynamic uncertainty relation (TUR) $\mathcal{Q}_\alpha\geq2$~\cite{Barato2015}. In terms of the affinity, violating the bound would require
\begin{align}
    \mathcal{F}_h<\mathcal{F}_h^{\rm TUR}
    =
    \frac{1}{\mathcal{A}},
    \qquad
    \mathcal{F}_c<\mathcal{F}_c^{\rm TUR}
    =
    \frac{2}{\mathcal{A}}.
    \label{eq:qhe-fano-tur-thresholds}
\end{align}
The Fano thresholds in Eq.~\eqref{eq:qhe-fano-tur-thresholds} are the quantities shown in Figs.~\ref{fig:qhe-antibunching} and~\ref{fig:qhe-finite-affinity-tur}. Note that the TUR can be saturated in the trivial case where $\mathcal{A} = 0$, since $\sigma = 0$; below, we concentrate in parameter regimes away from the reversible scenario, i.e., we focus always on finite-affinity regimes.

To diagnose whether the resonant non-linear process is active in the steady state, we also monitor the steady-state coherence
\begin{align}
    \mathcal{C}_{\rm RWA}
    =
    \sum_{n_h=1}^{N_h}
    \sum_{n_c=0}^{N_c-2}
    \left|
    \bra{n_h-1,n_c+2}\rho_{\rm ss}\ket{n_h,n_c}
    \right| .
    \label{eq:qhe-crwa}
\end{align}
This quantity measures steady-state coherence on the same links that enter the RWA heat-engine process. It is not a TUR bound, and it should not be interpreted as an optimisation target by itself. This distinction is consistent with recent work on quantum and thermodynamic uncertainty relations; coherence can enable reductions of current noise but does not, by its mere presence or magnitude, guarantee a thermodynamic advantage or a TUR violation~\cite{AgarwallaSegal2018,Prech2023,AlmanzaMarreroManzano2025}.

\subsection{Antibunching without TUR violation}

We first choose a large-affinity regime, $\bar n_h=0.5$ and $\bar n_c=0.01$, with $\lambda_h=0.47$, $\lambda_c=0.89$, $\kappa_h=2$, and $\kappa_c=0.5$. The frequency ratio is fixed to $\Omega_h/\Omega_c=\pi$ and $\Omega_c=1000$, which suppresses low-order competing resonances and keeps the near-resonant correction small, as discussed in Appendix~\ref{app:qhe-rwa}. Sweeping $g$ identifies the strongest hot-current antibunching near $g\simeq7.21$. A complementary sweep in $\lambda_c$ at this coupling gives the strongest antibunching near $\lambda_c\simeq0.85$, while the surrounding $\lambda_c\simeq0.83$--$0.89$ window remains close to the optimum. This reduction of current noise is the blockade-like effect produced by the Laguerre matrix elements.

The hot Fano factor is clearly sub-Poissonian in this regime, reaching $\mathcal{F}_h\simeq0.59$. This is a genuine quantum-optical signature of the incoming heat current. However, the large affinity, $\mathcal{A}\simeq4.07$, pushes the TUR threshold down to $\mathcal{F}_h^{\rm TUR}\simeq0.246$, and the corresponding uncertainty product remains around $\mathcal{Q}_h\simeq4.82$. 

This highlights that antibunching alone does not imply a thermodynamic precision advantage, and, in fact, is generally unrelated to it. Although antibunching is a genuine quantum effect \textit{for bosons}—i.e., their particle-like behaviour—this does not mean that antibunching cannot be understood via a classical rate-equation model~\cite{Kalee2021, Prech2023, Janovitch2023}.

\subsection{Saturating the TUR bound at finite affinity}

\begin{figure}[t]
    \centering
    \includegraphics[width=\linewidth]{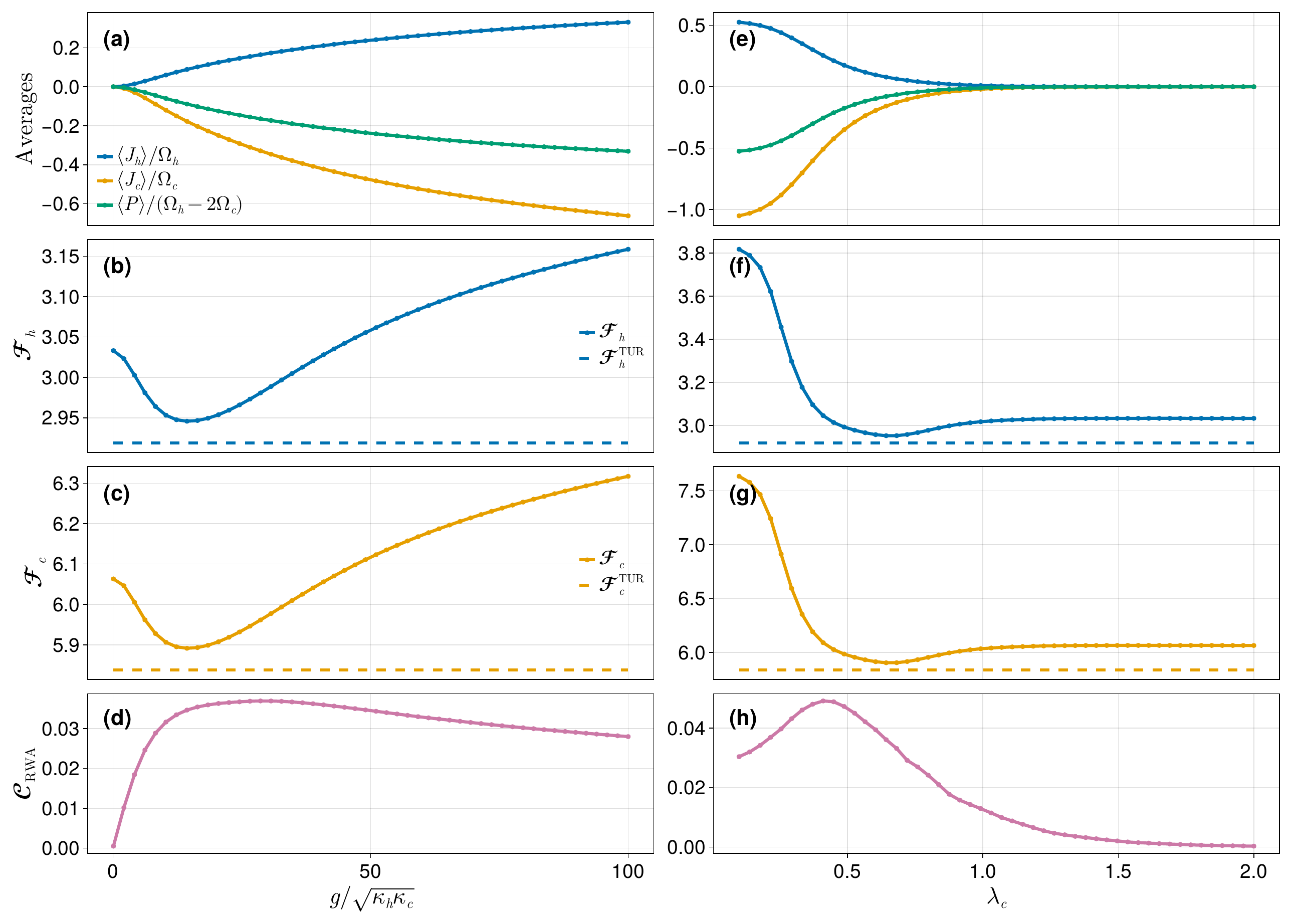}
    \captionof{figure}{
    Finite-affinity regime approaching the TUR bound. The left column shows the $g$ sweep at fixed $\lambda_c=0.7$, $\bar n_h=2.5$, and $\bar n_c=1.5$, while the right column shows the $\lambda_c$ sweep at fixed $g=10$. Panels \textbf{(a)} and \textbf{(e)} show the scaled heat currents and output power, \textbf{(b)} and \textbf{(f)} the hot Fano factor with its TUR threshold, \textbf{(c)} and \textbf{(g)} the corresponding cold Fano factor, and \textbf{(d)} and \textbf{(h)} the RWA-link coherence $\mathcal{C}_{\rm RWA}$. Reducing the thermodynamic affinity raises the Fano-factor thresholds and brings the hot and cold uncertainty products close to the classical bound while the currents and output power remain small.
    }
    \label{fig:qhe-finite-affinity-tur}
\end{figure}

The second regime focuses on attempting TUR violations at finite-affinity. We reduce the affinity by increasing the bath occupations to $\bar n_h=2.5$ and $\bar n_c=1.5$, giving $\mathcal{A}\simeq0.343$ for each elementary $(1,2)$ conversion event. The scans keep $\lambda_h=0.47$, $\kappa_h=2$, $\kappa_c=0.5$, $\Omega_c=1000$, and $\Omega_h/\Omega_c=\pi$. The left column of Fig.~\ref{fig:qhe-finite-affinity-tur} sweeps $g$ at fixed $\lambda_c=0.7$, while the right column sweeps $\lambda_c$ at fixed $g=10$. These simulations use the larger Hilbert-space cutoffs $N_{\max,h}=20$ and $N_{\max,c}=25$, with the numerical tolerances documented in Appendix~\ref{app:qhe-rwa}.

The corresponding Fano thresholds are much higher than in the antibunching scan, $\mathcal{F}_h^{\rm TUR}\simeq2.92$ and $\mathcal{F}_c^{\rm TUR}\simeq5.84$. The hot and cold Fano factors approach these thresholds from above; the best points in the displayed cuts occur near $g\simeq10.29$ and $\lambda_c\simeq0.72$, with $\mathcal{Q}_h\simeq\mathcal{Q}_c\simeq2.018$. The approach to the bound is accompanied by mild heat currents and output power, and a peak in $\mathcal{C}_\text{RWA}$.

The comparison between Figs.~\ref{fig:qhe-antibunching} and~\ref{fig:qhe-finite-affinity-tur} separates two effects. Laguerre-blockade physics can make the hot current sub-Poissonian at large affinity, but this remains far above the TUR threshold. Lowering the affinity moves the uncertainty products close to $\mathcal{Q}=2$, and the peak in coherence suggests that it is relevant in saturating the TUR bound at finite affinity, although it does not suppress fluctuations enough for a TUR violation. Within the checked RWA-compatible parameter windows, the non-linear heat engine therefore shows antibunching and coherence-assisted noise reduction without a finite-affinity TUR violation.

Truncation checks and RWA-validity diagnostics are provided in the Appendix~\ref{app:qhe-rwa} and in the accompanying reproducible notebooks.
Details on the thermodynamic identities and rotating-wave approximation are collected in Appendix~\ref{app:circuit-qhe}.

\section{Benchmarking}
\label{sec:benchmarking}

The recursive computation of full-counting statistics requires repeated applications of
the Drazin inverse of the Liouvillian. The numerical cost of this step determines which
Hilbert-space dimensions can be treated in practice, especially when the system contains
bosonic modes whose Fock spaces must be truncated. We therefore benchmark
\texttt{QuantumFCS.jl} using its default \code{method=:lu} against \texttt{Melt}~\cite{Melt}, a Mathematica package which includes tools for computing the first and second cumulants in FCS \cite{Melt}. As noted in Sec.~\ref{sec:introduction}, \texttt{Melt} is the only publicly available tool we are aware of that directly computes these quantities, and it is restricted to the first two cumulants. Note that FCS can also be computed by solving the time-evolution of the open quantum system and relying on the quantum regression theorem to compute and integrate correlation functions—this method is highly inefficient and will not be included in our benchmarking.

The comparison is restricted to the first two cumulants, $c_1$ and $c_2$, because
\texttt{Melt} can only compute the average current and noise. By contrast,
\texttt{QuantumFCS.jl} uses the same function call to compute cumulants up to
arbitrarily high order $n_C$, limited only by numerical cost and conditioning. In the
benchmarks below we set $n_C=2$ in order to compare both packages on the common task.
All timings measure the cumulant evaluation after construction of the model operators and
computation of the steady state. For \texttt{method = :lu}, the package automatically dispatches to the dense or sparse workflow according to the representation of the supplied objects, which is preserved by the \texttt{QuantumOptics.jl} and \texttt{QuantumToolbox.jl} interfaces.

We first consider the linearised quantum heat-engine model
\begin{equation}
    H_{\rm lin} = g(a_c^\dagger a_h + a_h^\dagger a_c),
    \label{eq:benchmark-linearised-ham}
\end{equation}
obtained from setting $l=k=1$ in Eq.~\eqref{eq:qhe-general-rwa}.
The monitored current is the cold-bath particle current, specified by the monitored jumps
$(J_2,J_4)$ and weights $(-1,1)$. The local Fock cutoff is increased symmetrically in the
hot and cold modes, so that the total Hilbert-space dimension is $d=N_hN_c$. This benchmark
tests the sparse-operator path used for weakly non-linear and linear transport
models.

\begin{figure}[t]
    \centering
    \includegraphics[width=\textwidth]{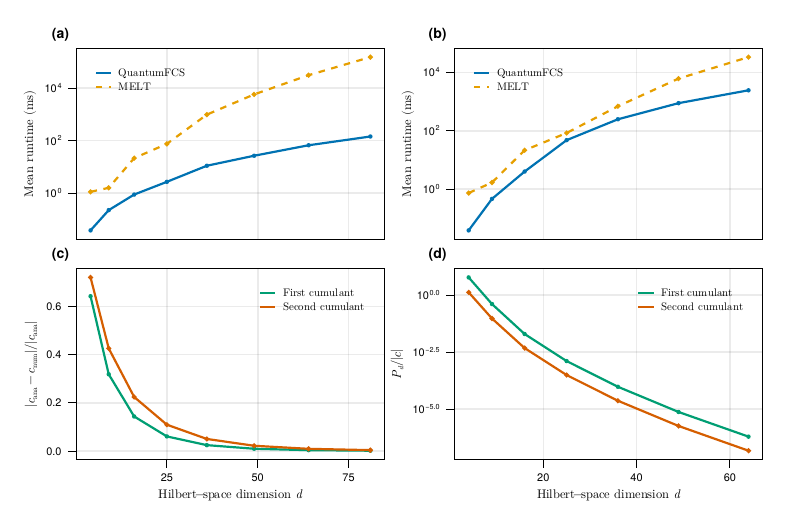}
    \caption{Benchmark summary for the linearised and dense quantum heat-engine
    models. \textbf{(a)} performance comparison for the linearised quantum heat-engine
    model, showing the mean runtime required to compute the first two cumulants as a
    function of Hilbert-space dimension using the default LU method. \texttt{QuantumFCS.jl} remains faster than
    \texttt{Melt} throughout the tested range, with the largest speed-ups appearing at
    the largest cutoffs. \textbf{(b)} performance comparison for the dense zero-bias
    circuit-QED heat-engine benchmark again using the default LU method. The dense benchmark is more demanding for both
    packages, but \texttt{QuantumFCS.jl} remains faster across all tested cutoffs.
    \textbf{(c)} relative error between the analytical result $c_{\rm ana}$ and the numerical result $c_{\rm num}$ of the first two cumulants in the linearised model, showing convergence with increasing Hilbert
    space. \textbf{(d)} highest retained joint Fock-state population, scaled by the
    magnitude of each dense cumulant, providing a cutoff-convergence diagnostic for the
    dense benchmark.}
    \label{fig:benchmark-linearised}
    \label{fig:benchmark-dense}
\end{figure}

As shown in Fig.~\ref{fig:benchmark-linearised}(a), the sparse benchmark displays a
substantial performance advantage for \texttt{QuantumFCS.jl}. The observed speed-up grows from roughly one order
of magnitude at small dimensions to more than three orders of magnitude at the largest
tested dimension. Panel (c) shows the relative error between the analytical cumulants $c_{\rm ana}$ from Refs.~\cite{Kerremans2022,Janovitch2023} and the numerical results $c_{\rm num}$. Classical models whose dynamics remains diagonal are likewise routed automatically through this workflow when supplied in a sparse representation and are therefore expected to benefit from the same performance advantage.

We next benchmark the dense circuit-QED heat-engine model
\begin{equation}
    H = \Omega_h a_h^\dagger a_h + \Omega_c a_c^\dagger a_c
        - E_J \cos(2\Phi),
    \qquad
    \Phi = \lambda_h(a_h^\dagger+a_h) + \lambda_c(a_c^\dagger+a_c),
    \label{eq:benchmark-dense-ham}
\end{equation}
which is obtained from Eq.~\eqref{eq:qhe-lab-hamiltonian} in the zero-bias regime, i.e., setting $V=0$.
This model retains the full cosine interaction in the truncated Fock basis and therefore
requires dense matrix operations. The monitored current is the hot-channel current,
specified by
\begin{equation}
    \mathbf{mJ} = (J_1,~J_3),~\boldsymbol{\nu} = (-1,~1).
\end{equation}
% monitored jumps $(J_1,J_3)$ and weights $(-1,1)$.
This case complements the
linearised benchmark by testing a non-linear model whose operator representation is dense.

Figure~\ref{fig:benchmark-dense}(b) shows that the advantage persists in the dense setting.
The speed-up is more moderate than in the sparse benchmark,
but remains favourable over the full range, varying from about $1.8$ to $19$.

The benchmarks were performed on a workstation with a 12th Gen Intel(R) Core(TM)
i7-12800HX CPU, using one Julia thread. The software environment used Julia 1.12.6,
\texttt{QuantumFCS.jl} v1.1.0, and Mathematica 14.3.0.0 for
\texttt{Melt}. Additional model definitions, parameter tables, and reproducibility
details are collected in Appendix~\ref{app:benchmarking-details}. These results demonstrate that the recursive implementation in
\texttt{QuantumFCS.jl} provides both the flexibility to compute higher-order cumulants
and a substantial runtime advantage for the benchmarked first- and second-cumulant
calculations. We emphasise that these are cross-language comparisons: the measured
speed-ups reflect both algorithmic differences and language- and implementation-level
factors between a compiled Julia package and a Mathematica implementation. They should
therefore be read as a practical performance comparison against the only available tool,
rather than as an isolated measurement of algorithmic complexity.

The scripts, processed benchmark data, and plotting routines used to generate
Fig.~\ref{fig:benchmark-dense} are available in the accompanying repository~\cite{QuantumFCSNotebooks}.

\section{Summary and outlook}
\label{sec:summary}

In this work we introduced \texttt{QuantumFCS.jl}, an open-source Julia package for computing zero-frequency full-counting statistics of currents in quantum systems described by Lindblad master equations. The package is built around three practical design choices: currents are specified through monitored jump operators and weights, cumulants are generated from recursive relations without numerical differentiation of the tilted-Liouvillian eigenvalue, and dense or sparse solver backends can be selected from the same problem interface.

We described both the problem-object interface and the direct function interface of the package. The implementation supports dense and sparse Liouvillian representations, as well as models constructed with high-level quantum-optics packages or with raw matrix inputs. For large sparse problems, the Drazin action required in the recursion can be evaluated through gauge-fixed linear solves, allowing iterative Krylov methods and preconditioning to be used. The minimal quantum-dot example illustrates the basic workflow, recovers the analytical first three ordinary cumulants of the transport current, and shows how the same calculation returns factorial cumulants for dimensionless event counts.

We then applied the package to two more demanding examples. For the driven-dissipative Jaynes-Cummings model, the iterative sparse backend makes it possible to compute photon-current cumulants in the blockade-breakdown regime, where the Fano factor and the third cumulant are sensitive to the crossover between dim and bright emission. For the circuit-QED heat engine, the package provides direct access to heat-current fluctuations in a non-linear Cooper-pair-assisted conversion process. This allows us to distinguish current antibunching from thermodynamic-uncertainty reduction and to identify finite-affinity regimes where the uncertainty product approaches, but does not violate, the classical thermodynamic uncertainty bound at finite affinity.

The benchmarks against \texttt{Melt} show that \texttt{QuantumFCS.jl} provides consistent first- and second-cumulant results while giving substantial runtime improvements in the tested sparse and dense heat-engine models. They also highlight a practical advantage of the recursive implementation: the same workflow can be used to compute higher-order cumulants, limited mainly by numerical cost and conditioning.

The current implementation focuses on long-time, or zero-frequency, current cumulants of quantum Markovian dynamics, and we implement an efficient method for large sparse matrices (iLU + GMRES).
Natural directions in which we plan to extend \texttt{QuantumFCS.jl} are:
\begin{enumerate}
    \item[\textcolor{quantumviolet}{E1}] \textbf{Finite-time and finite-frequency FCS.} Long-time cumulants are not adequate characterisations of full-counting statistics for transient dynamics or genuinely time-dependent systems. In these situations, the recursive scheme on which our implementation is based~\cite{Flindt2008} does not apply. Instead, FCS has to be computed by solving the time-evolution generalised master equations directly; for example, either the count-resolved density matrix on the right-hand side of Eq.~\eqref{eq:dressed-state} or the counting-field-resolved density matrix on the left-hand side of Eq.~\eqref{eq:dressed-state}. The Julia ecosystem provides a set of powerful differential-equation solvers \cite{rackauckas2017differentialequations} that can be used for this implementation. Unlike the recursive method, however, this approach implies a significant increase in numerical complexity, since for each count or counting-field value an object of the size of the original density matrix has to be evolved.
    \item[\textcolor{quantumviolet}{E2}] \textbf{Non-Markovian FCS.} The recursive scheme of Refs.~\cite{Flindt2008, Flindt2010} is to a great extent already applicable to non-Markovian systems. Recent progress in compactly representing non-Markovian kernels with process tensors~\cite{Strathearn2018, Cygorek2022, Keeling2026} provides clear implementation pathways. Non-Markovian FCS has been computed using these and further methods~\cite{Popovic2021, Shubrook2025, Valli2025, Diba2024}, but, to the best of our knowledge, no general-purpose implementation is publicly available. This topic has gained much attention, as non-Markovian FCS has been used as a tool to characterise transport in closed many-body systems~\cite{Valli2025} and directly probed in quantum computers~\cite{Google2024}.
    \item[\textcolor{quantumviolet}{E3}] \textbf{Expanding solver backends.} The iterative method based on the iLU + GMRES implementation focused on improving memory usage to enable tabletop calculations of large quantum systems described by sparse Liouvillians. Different computer architectures such as dedicated clusters, or different classes of sparse problems may benefit from exploring workflows based on other sparse solvers. A wealth of methods is available in \texttt{Krylov.jl}~\cite{Montoison2023}.
\end{enumerate}

\section*{Code and data availability}

\texttt{QuantumFCS.jl} is free and open-source software, released under the MIT licence and
registered in the Julia General registry; its source code is available at
\url{https://github.com/marcelojbp/QuantumFCS}~\cite{QuantumFCS}. The scripts, processed
data, and plotting routines that generate all figures in this manuscript---together with
the minimal quantum-dot, driven-dissipative Jaynes-Cummings, and circuit-QED heat-engine
workflows---are collected in the companion repository~\cite{QuantumFCSNotebooks}. The
benchmark comparison against \texttt{Melt} additionally provides the raw and processed
timing data used in Sec.~\ref{sec:benchmarking}.

\section*{Author contributions}
Both authors contributed equally in developing the software, analysing relevant models, and writing the manuscript. M.J. conceived the original idea and led the project.

\section*{Acknowledgements}
The authors thank the organisers of the SimTech school on Research Software Engineering (Stuttgart, 2023), where this project was born. The authors acknowledge valuable discussion with Patrick P. Potts, Mark T. Mitchison, Alberto Mercurio, Kacper Prech, and Matteo Brunelli. A. D. acknowledges funding from the QCQT PhD School.
\bibliographystyle{quantum}

\onecolumn
\appendix

\section{Numerical details for the driven-dissipative Jaynes-Cummings model}
\label{app:ddjc-numerics}

\subsection{Cutoff selection}
\label{app:ddjc-cutoff}

The driven-dissipative Jaynes-Cummings sweep in
Fig.~\ref{fig:ddjc-fcs-scans} uses a fixed coupling $g/\kappa=14$ and varies
$2\mathcal{E}/g$ at fixed detuning. Ref.~\cite{Carmichael2015BreakdownPhotonBlockade}
uses the cavity-loss convention
$\kappa_C(2a\rho a^\dagger-a^\dagger a\rho-\rho a^\dagger a)$, whereas
Eq.~\eqref{eq:ddjc-master-equation} uses $\kappa\mathcal{D}[a]\rho$. Thus the
linewidths are related by $\kappa=2\kappa_C $
% \end{align}
With this conversion, we use the semi-classical equations of
Ref.~\cite{Carmichael2015BreakdownPhotonBlockade} only as a bright-branch
photon-number estimate for choosing the Fock cutoff before solving the FCS problem. On resonance, the above-threshold branch gives
\begin{align}
    n_{\rm sc}(\Delta=0)
    =
    \max\left[
    \frac{4\mathcal{E}^2-g^2}{\kappa^2},
    0
    \right],
    \label{eq:ddjc-resonant-cutoff-estimate}
\end{align}
so the critical drive $2\mathcal{E}/g=1$ corresponds to the natural scale
$n_{\rm scale}=(g/\kappa)^2=196$ for the sweep in
Fig.~\ref{fig:ddjc-fcs-scans}.

Away from resonance, the upper and lower ladders separate with local transition
detuning
\begin{align}
    E_{n+1,\pm}-E_{n,\pm}-\omega_D
    \simeq
    -\Delta\pm\frac{g}{2\sqrt{n}}~,
    \label{eq:ddjc-large-n-detuning}
\end{align}
a large-$n$ mean-field estimate. This large-$n$ ladder condition identifies the
positive-detuning channel
and gives the photon-number scale on which the bright branch first becomes
accessible. The estimate used for the cutoff is the largest positive root,
over the two dressed branches $s=\pm1$, of
\begin{align}
    n_{\rm sc}^{(s)}
    =
    \frac{\mathcal{E}^2}
    {(\kappa/2)^2+
    \left[
        \Delta
        -s\,{\rm sgn}(\Delta)
        \frac{g^2}{\sqrt{\Delta^2+4g^2 n_{\rm sc}^{(s)}}}
    \right]^2}.
    \label{eq:ddjc-detuned-cutoff-estimate}
\end{align}
Equation~\eqref{eq:ddjc-detuned-cutoff-estimate} is not used as a replacement
for the quantum steady state. It is a conservative pre-solve estimate of the
bright-branch occupation, used only to decide how large the truncated Fock
space must be.

The cutoff is therefore chosen point by point rather than fixed globally. The
available cutoff tiers are
\[
    N\in\{150,175,200,225,250,275,300,350,400,450,500\}.
\]
For each drive and detuning, a semi-classical bright-branch estimate sets a
target occupation $n_{\rm sc}$. The target cutoff is
\begin{align}
    N_{\rm target}
    =
    \max\left[
        \frac{n_{\rm sc}}{\eta},
        n_{\rm sc}+p_\sigma\sqrt{n_{\rm sc}}+p_0
    \right],
    \qquad
    \eta=0.5,\quad p_0=25,
    \label{eq:ddjc-cutoff-target}
\end{align}
with $p_\sigma=14$ on resonance and $p_\sigma=6$ for the detuned cuts.

The two arguments of the maximum encode different requirements. The second,
$n_{\rm sc}+p_\sigma\sqrt{n_{\rm sc}}+p_0$, is a resolution requirement: on the
bright branch the cavity state is coherent-like, with a Fock-space width
$\approx\sqrt{n_{\rm sc}}$, so $p_\sigma$ counts the standard deviations of
headroom retained above the estimated mean occupation, while the absolute floor
$p_0$ keeps the cutoff sensible in the blockaded region, where $n_{\rm sc}$, and
hence $\sqrt{n_{\rm sc}}$, are small. This is the criterion that decides whether
a point is genuinely under-resolved; at $p_\sigma=6$ it leaves a boundary
population of order $10^{-9}$. The first argument, $n_{\rm sc}/\eta$, instead
caps the mean occupation at a fraction $\eta$ of the cutoff. It never flags a
point: it only raises the selected tier, and it acts as a conservative backstop
in the coexistence region, where the semi-classical estimate over-reports the
quantum occupation.

The additional headroom is required by the third cumulant. While $c_1$ and
$\mathcal{F}=c_2/c_1$ converge with modest headroom, $c_3$ is considerably more
sensitive to the retained tail. The quantity controlling this sensitivity is the
relative headroom
$(N-\langle a^\dagger a\rangle_{\rm ss})/\sqrt{\langle a^\dagger a\rangle_{\rm ss}}$,
which at fixed occupation fraction grows only as $\sqrt{N}$. The resonant cut
runs at the smallest photon numbers and therefore has the least relative
headroom at a given tier; it is assigned $p_\sigma=14$ rather than the
$p_\sigma=6$ used on the detuned cuts, which raises its bright-branch headroom
from $\approx10$ to $\approx16$ standard deviations---the level at which $c_3$
is converged---while leaving the cutoffs selected on the detuned cuts unchanged.
The comparatively tight occupation cap $\eta=0.5$ serves the same purpose. The
tier spacing (25 photons up to $N=300$, 50 above) is likewise chosen so that the
cutoff changes only mildly between neighbouring drive points.

The
selected $N$ is the smallest available tier above
Eq.~\eqref{eq:ddjc-cutoff-target}, capped at $N=500$.
This rule avoids using
an unnecessarily large Hilbert space in the blockaded part of the sweep while
still resolving the bright branch. The actual quantum solution is then checked
by the boundary population $p_{N-1}+p_N$ and the relative occupation $\langle a^\dagger a\rangle/N$ rather than trusted solely from the
semi-classical estimate, and recomputed with different larger $N$ if necessary for further confidence.

\subsection{Preconditioned sparse solves}
\label{app:ddjc-preconditioner}

At cutoff $N$, the Hilbert-space dimension is $2(N+1)$ and the vectorised
Liouville-space dimension is $[2(N+1)]^2$. For the largest cutoff used in
Fig.~\ref{fig:ddjc-fcs-scans}, this is $1004^2$. The Liouvillian is sparse and
non-Hermitian, but sparse direct factorisations still suffer from fill-in (excessive memory usage) at
these dimensions. As in
Sec.~\ref{sec:sparse-iterative-backend}, the steady state is obtained from a
trace-constrained GMRES solve with a shifted-iLU preconditioner.

The sweep is ordered so that neighbouring drive points at fixed detuning are
solved consecutively. Within a fixed cutoff tier, the previous steady state is
used as the initial GMRES vector. The incomplete-LU factorisation is also
reused until the iteration count indicates that the local linear problem has
changed enough to justify rebuilding it. This continuation strategy is
particularly effective because the expensive step is often constructing the
preconditioner, not applying GMRES once a useful preconditioner is available. 

The same preconditioner then serves as the right
preconditioner for the Drazin solves, exactly as in
Sec.~\ref{sec:sparse-iterative-backend}.

The full simulations in Fig.~\ref{fig:ddjc-fcs-scans} for the three detuning sweeps took about
$45\,{\rm min}$ on the MacBook Pro M1 with 16 GB RAM noted in
Sec.~\ref{sec:ddjc}. The machine was kept otherwise quiet during the run:
the $N=500$ points get close enough to the available memory that competing
workloads can trigger swapping and make the timings unreliable.

\subsection{Numerical validation}
\label{app:ddjc-validation}

Figure~\ref{fig:ddjc-fcs-scans} includes three diagnostics alongside the
physical cumulants. The first is the selected cutoff tier, which reveals where
the Hilbert space grows along each sweep. The second is the boundary population
$p_{N-1}+p_N$, compared with the threshold
$\epsilon_{\rm trunc}=10^{-5}$. In the final sweep the largest boundary
population is below $5\times10^{-6}$. The third is
the occupation fraction
$\langle a^\dagger a\rangle_{\rm ss}/N$, which tests whether the distribution
is approaching the edge of the retained Fock space even when the boundary tail
is small; for the latter, we checked that 0.7 is still a conservative estimate since the 
Fock-state population distribution is not heavy tailed above the average photon number.

We also use two checks that are not separately plotted. First, the mean count
obtained from FCS must agree with the steady-state cavity-emission identity
$c_1=\kappa\langle a^\dagger a\rangle_{\rm ss}$. The largest relative mismatch
in the final data is below $7\times10^{-8}$. Second, selected points near the
largest occupations and the sharp detuned jumps are recomputed one cut-off tier
lower. This directly checks the sensitivity of the cumulants, including the
third cumulant, to the truncation. A few points where the semi-classical hard
cutoff estimate reaches beyond the $N=500$, or where $\langle a^\dagger a\rangle_{\rm ss}/N > 0.5$ were recalculated with $N=600, 700$, showing no relevant change in the Fock state distribution and FCS.

\section{Details for the circuit-QED heat-engine application}
\label{app:circuit-qhe}

\subsection{Thermodynamic Hamiltonian and current identities}
\label{app:qhe-thermo}
The local thermal dissipator used in Sec.~\ref{sec:circuit-qhe} is
\begin{align}
    \dv{\rho}{t}
    &=
    -i[H(t),\rho]
    + \sum_{\alpha=h,c}\mathcal{L}_\alpha\rho,
    \nonumber\\
    \mathcal{L}_\alpha\rho
    &=
    \kappa_\alpha(\bar n_\alpha+1)\mathcal{D}[a_\alpha]\rho
    + \kappa_\alpha \bar n_\alpha\mathcal{D}[a_\alpha^\dagger]\rho ,
    \label{eq:qhe-lab-lme}
\end{align}
where $\bar n_\alpha$ is the Bose-Einstein occupation of bath $\alpha$. For the RWA model selected in the main text, the thermodynamic Hamiltonian is
\begin{align}
    H_{\rm TD} = \Omega_h a_h^\dagger a_h + \Omega_c a_c^\dagger a_c .
    \label{eq:qhe-thermodynamic-hamiltonian}
\end{align}
The corresponding heat current associated with bath $\alpha$ is
\begin{align}
    \langle J_\alpha\rangle
    =
    \tr\left\{H_{\rm TD}\mathcal{L}_\alpha\rho_{\rm ss}\right\}
    =
    \kappa_\alpha\Omega_\alpha\left(\bar n_\alpha-\langle a_\alpha^\dagger a_\alpha\rangle_{\rm ss}\right).
    \label{eq:qhe-heat-current}
\end{align}
With the convention that positive heat current flows into the system, the entropy production rate is
\begin{align}
    \sigma
    &=
    -\frac{\langle J_h\rangle}{T_h}
    -\frac{\langle J_c\rangle}{T_c} \nonumber\\
    &=
    \frac{\langle J_c\rangle}{\Omega_c}
    \left[
    \frac{1}{2}\ln(1+\bar n_h^{-1})
    -
    \ln(1+\bar n_c^{-1})
    \right],
    \label{eq:qhe-entropy-production}
\end{align}
where $\Omega_\alpha/T_\alpha=\ln(1+\bar n_\alpha^{-1})$ and the stationary current relation derived below has been used.

For the non-linear process in Eq.~\eqref{eq:qhe-non-linear-hamiltonian}, the interaction in the laboratory frame can be written as
\begin{align}
    H_{\rm nl}(t)
    =
    \frac{E_J}{2}
    \left[
    i(a_c^\dagger)^2 A_c(2)A_h(1)a_h e^{2ieVt}
    + \mathrm{H.c.}
    \right].
\end{align}
The power operator follows from $\partial_t H_{\rm nl}(t)$. In the rotating frame,
\begin{align}
    P
    =
    -eVE_J
    \left[
    (a_c^\dagger)^2A_c(2)A_h(1)a_h
    +
    a_h^\dagger A_h(1)A_c(2)a_c^2
    \right].
\end{align}
Defining the Cooper-pair current through $P=2eVI$, one obtains
\begin{align}
    I
    =
    -\frac{E_J}{2}
    \left[
    (a_c^\dagger)^2A_c(2)A_h(1)a_h
    +
    a_h^\dagger A_h(1)A_c(2)a_c^2
    \right].
\end{align}
Since $A_h(1)A_c(2)$ is diagonal in the number basis, it commutes with both $N_h=a_h^\dagger a_h$ and $N_c=a_c^\dagger a_c$. Therefore
\begin{align}
    -i[H_{\rm TD},H_{\rm nl}]
    =
    (\Omega_h-2\Omega_c)I .
\end{align}
Using the resonance condition $2eV=\Omega_h-2\Omega_c$ verifies that the bare oscillator Hamiltonian in Eq.~\eqref{eq:qhe-thermodynamic-hamiltonian} is the thermodynamic Hamiltonian for the RWA model.

The heat-current relation in Eq.~\eqref{eq:qhe-current-relation} follows from the conservation of $2N_h+N_c$ under the non-linear interaction. At stationarity,
\begin{align}
    \langle\mathcal{L}_\alpha^\dagger N_\alpha\rangle_{\rm ss}
    =
    -i\langle[H_{\rm nl},N_\alpha]\rangle_{\rm ss}.
\end{align}
Since $[H_{\rm nl},2N_h+N_c]=0$, the dissipative contributions satisfy
\begin{align}
    \langle\mathcal{L}_h^\dagger N_h\rangle_{\rm ss}
    =
    -\frac{1}{2}
    \langle\mathcal{L}_c^\dagger N_c\rangle_{\rm ss},
\end{align}
which gives the tight-coupling relation
\begin{align}
    2\Omega_c\langle J_h\rangle
    =
    -\Omega_h\langle J_c\rangle
    \label{eq:qhe-current-relation}
\end{align}
after multiplying by the corresponding mode energies.

\subsection{Rotating-wave consistency checks and numeric reliability}
\label{app:qhe-rwa}
% The general rotating-wave Hamiltonian for a process satisfying the resonance condition
% $2eV=k\Omega_h-l\Omega_c$ is
% \begin{align}
%     H_{k,l}
%     =
%     -\frac{E_J}{2}\left[
%     i^{k+l}(a_c^\dagger)^l A_c(l)A_h(k)a_h^k+\mathrm{H.c.}
%     \right],
%     \label{eq:qhe-general-rwa}
% \end{align}
% where the Laguerre matrix elements
% \begin{align}
%     A_\alpha(k)
%     &=
%     (2\lambda_\alpha)^k e^{-2\lambda_\alpha^2}
%     \sum_{n=0}^{\infty}
%     \frac{n!}{(n+k)!}
%     L_n^{(k)}(4\lambda_\alpha^2)\ket{n}\!\bra{n}
%     \label{eq:qhe-laguerre-operator}
% \end{align}
% are diagonal in the Fock basis, with $L_n^{(k)}$ the generalised Laguerre polynomial; in the
% numerical implementation this diagonal operator is built directly in the truncated Fock basis.

The selected process is isolated by choosing the voltage according to Eq.~\eqref{eq:qhe-non-linear-voltage}. Other resonant processes would solve
\begin{align}
    \Omega_h-2\Omega_c = k\Omega_h-l\Omega_c .
\end{align}
Writing $\Omega_h=c\Omega_c$ gives
\begin{align}
    c = \frac{2-l}{1-k}.
\end{align}
Thus additional exact resonances require a rational frequency ratio. The numerical examples therefore use a ratio close to an irrational value, for instance $\Omega_h/\Omega_c\simeq \pi$, so that low-order competing resonances are absent.

Near-resonant processes are controlled by comparing their matrix elements with their detuning, following the supplement of Ref.~\cite{HoferSouquetClerk2016}. For $\Omega_h=\pi\Omega_c$, the closest low-order process has detuning
\begin{align}
    \delta = (\pi-3)\Omega_c .
\end{align}
The corresponding off-resonant term is written as $F e^{i\delta t}+\mathrm{H.c.}$ with
\begin{align}
    F = -\frac{E_J}{2} i a_c A_h(0)A_c(1).
\end{align}
A useful dimensionless estimate is
\begin{align}
    \epsilon_{\rm off}
    =
    \frac{1}{|\delta|}
    \sqrt{
    \left\langle
    FF^\dagger+F^\dagger F
    \right\rangle_{\rho_{\rm ss}}
    }.
\end{align}
The RWA is self-consistent when $\epsilon_{\rm off}\ll1$ and the stationary Fock-state populations remain small near the Hilbert-space cutoff used in the numerical calculation.

\begin{figure}[H]
    \centering
    \IfFileExists{figures/qhe_antibunching_validation.pdf}
    {\includegraphics[width=0.95\textwidth]{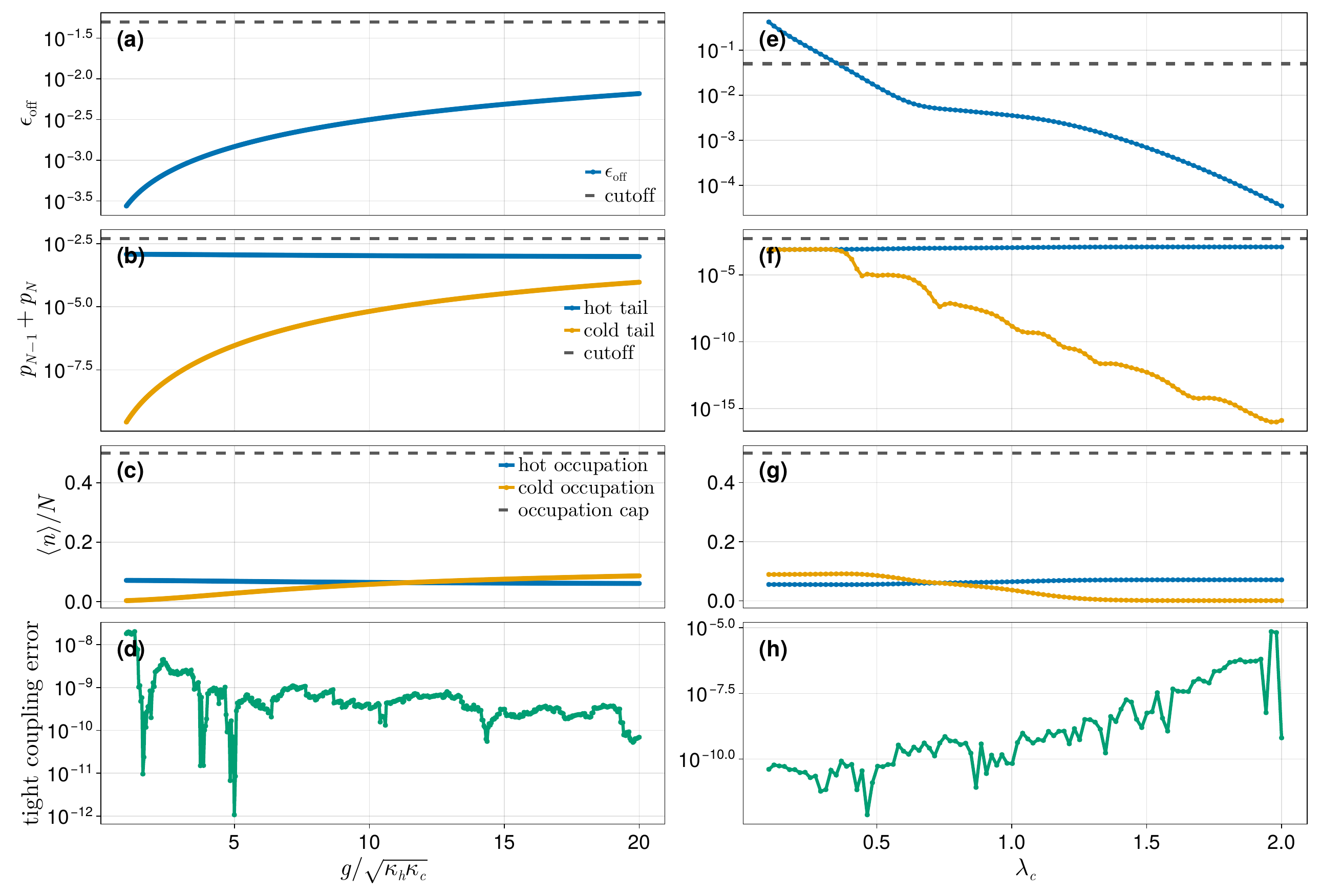}}
    {\fbox{\begin{minipage}[c][0.22\textheight][c]{0.9\textwidth}
    \centering
    Appendix diagnostic plot not yet generated.
    \end{minipage}}}
    \caption{
    Numerical diagnostics for the antibunching regime in Fig.~\ref{fig:qhe-antibunching}. The left column corresponds to the $g$ sweep and the right column to the $\lambda_c$ sweep. Panels \textbf{(a)} and \textbf{(e)} show the off-resonant estimate $\epsilon_{\rm off}$, \textbf{(b)} and \textbf{(f)} the hot and cold cutoff-tail probabilities, \textbf{(c)} and \textbf{(g)} the mean hot and cold occupations as fractions of the corresponding cutoff dimensions, and \textbf{(d)} and \textbf{(h)} the tight-coupling current-identity error. Dashed horizontal lines mark the thresholds used in the numerical filters.
    }
    \label{fig:qhe-antibunching-validation}
\end{figure}

\begin{figure}[H]
    \centering
    \IfFileExists{figures/qhe_finite_affinity_validation.pdf}
    {\includegraphics[width=0.95\textwidth]{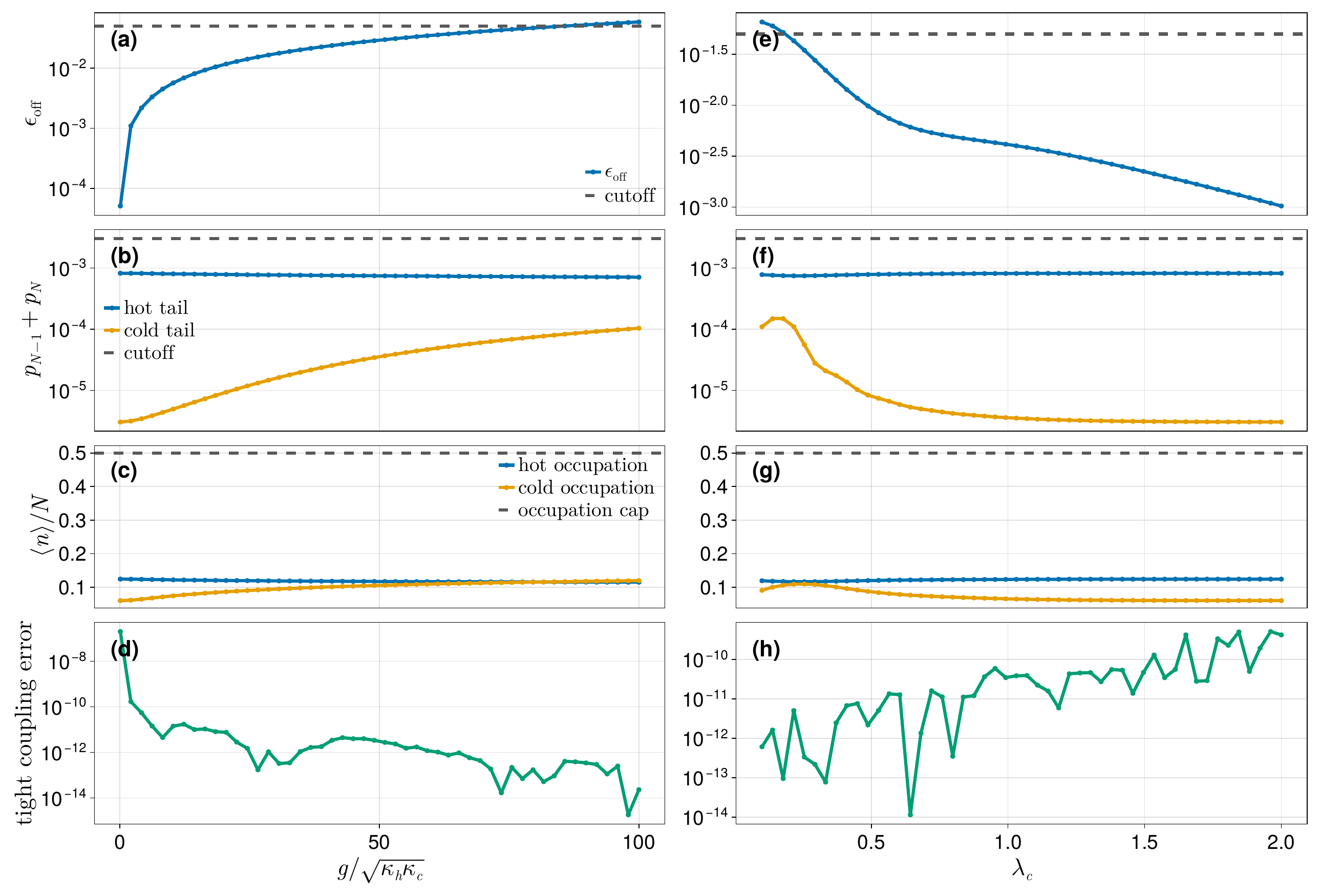}}
    {\fbox{\begin{minipage}[c][0.22\textheight][c]{0.9\textwidth}
    \centering
    Appendix diagnostic plot not yet generated.
    \end{minipage}}}
    \caption{
    Numerical diagnostics for the finite-affinity TUR regime in Fig.~\ref{fig:qhe-finite-affinity-tur}. The panel order matches Fig.~\ref{fig:qhe-antibunching-validation}: \textbf{(a)} and \textbf{(e)} show $\epsilon_{\rm off}$, \textbf{(b)} and \textbf{(f)} the cutoff-tail probabilities, \textbf{(c)} and \textbf{(g)} the occupation fractions, and \textbf{(d)} and \textbf{(h)} the tight-coupling current-identity error. The cutoff threshold is adjusted to the $3\times10^{-3}$ truncation tolerance used for this regime, and the occupation-fraction row is evaluated with $N_{\max,h}=20$ and $N_{\max,c}=25$.
    These checks verify that the approach of the uncertainty products towards the classical bound is not caused by visible cutoff leakage or by a breakdown of the selected RWA process.
    }
    \label{fig:qhe-finite-affinity-validation}
\end{figure}

For the parameter cuts shown in the main text we monitor these checks directly.
The first diagnostic is $\epsilon_{\rm off}$, which estimates the size of the closest neglected near-resonant contribution relative to its detuning.
The second is the sum of the two highest retained marginal Fock-state populations for each bosonic mode, which tests whether the Hilbert-space cutoff is reached.
The third is the mean occupation as a fraction of the chosen cutoff, which provides a complementary check that the displayed trends are not driven by the top of the retained Fock ladder.
The fourth is the residual violation of the tight-coupling current identity in Eq.~\eqref{eq:qhe-current-relation}.
Figures~\ref{fig:qhe-antibunching-validation} and~\ref{fig:qhe-finite-affinity-validation} show these quantities for the same one-dimensional cuts used in Figs.~\ref{fig:qhe-antibunching} and~\ref{fig:qhe-finite-affinity-tur}.
The antibunching diagnostics use $N_{\max,h}=7$, $N_{\max,c}=7$ for the $g$ sweep and $N_{\max,h}=7$, $N_{\max,c}=10$ for the $\lambda_c$ sweep, with tail tolerance $5\times10^{-3}$.
The finite-affinity diagnostics use $N_{\max,h}=20$ and $N_{\max,c}=25$, with tail tolerance $3\times10^{-3}$.
Together these checks support the use of the truncated RWA model in the regimes discussed in the main text.

\section{Additional benchmarking details}
\label{app:benchmarking-details}

This appendix records the model definitions, numerical parameters, and reproducibility
details used for the benchmarking results in the main text. The purpose is
to make the comparison between \texttt{QuantumFCS.jl} and \texttt{Melt} reproducible
without adding implementation-level detail to the main text.

\subsection{Linearised heat-engine benchmark}
\label{app:benchmarking-linearised}

The linearised benchmark uses the two-mode Hamiltonian of
Eq.~\eqref{eq:benchmark-linearised-ham}, with local thermal jump operators
\begin{equation}
    \mathbf{J} =
    \left(
        \sqrt{(\bar n_h+1)\kappa_h}\,a_h,\,
        \sqrt{(\bar n_c+1)\kappa_c}\,a_c,\,
        \sqrt{\bar n_h\kappa_h}\,a_h^\dagger,\,
        \sqrt{\bar n_c\kappa_c}\,a_c^\dagger
    \right).
\end{equation}
The monitored current is the cold-bath particle current, so the monitored jumps and
weights are
\begin{equation}
    \mathbf{mJ} = (J_2,J_4),
    \qquad
    \boldsymbol{\nu} = (-1,1).
\end{equation}
The parameters used in the benchmark are listed in Table~\ref{tab:linearised-benchmark-params}.

\begin{table}
    \centering
    \begin{tabular}{ll}
        \hline
        Parameter & Value \\
        \hline
        Coupling & $g=0.35$ \\
        Hot occupation & $\bar n_h=0.5$ \\
        Cold occupation & $\bar n_c=0.05$ \\
        Dissipation rates & $\kappa_h=\kappa_c=1.0$ \\
        Computed cumulants & $n_C=2$ \\
        Julia local cutoffs & $N=1,\ldots,8$ \\
        Julia local dimensions & $d=N+1=2,\ldots,9$ \\
        \texttt{Melt} local dimensions & $d=2,\ldots,9$ \\
        Julia samples per point & $100$ \\
        \texttt{Melt} samples per point & $5$ \\
        Julia evaluations per sample & $1$ \\
        \hline
    \end{tabular}
    \caption{Parameters for the linearised heat-engine benchmark. In the Julia
    implementation, \code{FockBasis(N)} contains the states $0,\ldots,N$, so the
    local Hilbert-space dimension is $d=N+1$.}
    \label{tab:linearised-benchmark-params}
\end{table}

For this model the first two cumulants can be compared with the analytic
expressions valid for the RWA Gaussian model~\cite{Janovitch2023}. With $\kappa_h=\kappa_c=\kappa$, the reference current is
\begin{equation}
    \langle I\rangle = \frac{2g^2\kappa}{4g^2+\kappa^2}(\bar n_h-\bar n_c),
\end{equation}
and the monitored cold-bath convention gives the first cumulant $c_1=-\langle I\rangle$. The second cumulant plotted in the convergence panel is compared against
\begin{equation}
    c_2 =
    E\left[\bar n_h(\bar n_h+1)+\bar n_c(\bar n_c+1)\right]
    - S(\bar n_h-\bar n_c)^2,
    \qquad
    E=\frac{\langle I \rangle}{\bar n_h-\bar n_c},
\end{equation}
where
\begin{equation}
    S =
    E\left[
        1-\frac{2g^2(4g^2+5\kappa^2)}{(4g^2+\kappa^2)^2}
    \right].
\end{equation}

\subsection{Dense zero-bias circuit-QED benchmark}
\label{app:benchmarking-dense}

The dense benchmark uses the zero-bias circuit-QED Hamiltonian of
Eq.~\eqref{eq:benchmark-dense-ham}.
The same local thermal dissipator structure is used, with occupations denoted
$\bar n_h$ and $\bar n_c$. The monitored current is the hot-channel current,
\begin{equation}
    \mathbf{mJ} = (J_1,J_3),
    \qquad
    \boldsymbol{\nu}=(-1,1).
\end{equation}
The parameters used in the dense benchmark are listed in
Table~\ref{tab:dense-benchmark-params}.

\begin{table}
    \centering
    \begin{tabular}{ll}
        \hline
        Parameter & Value \\
        \hline
        Mode frequencies & $\Omega_h=5.0$, $\Omega_c=1.0$ \\
        Josephson energy & $E_J=1.75$ \\
        Phase fluctuations & $\lambda_h=0.20$, $\lambda_c=0.25$ \\
        Dissipation rates & $\kappa_h=\kappa_c=1.0$ \\
        Thermal occupations & $\bar n_h=0.50$, $\bar n_c=0.05$ \\
        Computed cumulants & $n_C=2$ \\
        Julia local cutoffs & $N=1,\ldots,7$ \\
        Julia local dimensions & $d=N+1=2,\ldots,8$ \\
        \texttt{Melt} local dimensions & $d=2,\ldots,8$ \\
        Julia samples per point & $10$ \\
        \texttt{Melt} samples per point & $10$ \\
        Julia evaluations per sample & $1$ \\
        \hline
    \end{tabular}
    \caption{Parameters for the dense zero-bias circuit-QED heat-engine benchmark.}
    \label{tab:dense-benchmark-params}
\end{table}

The dense benchmark includes an additional cutoff diagnostic. For each truncation we
record the steady-state population of the joint cutoff-corner state
$|d_h-1,d_c-1\rangle$. In the summary figure this population is divided by the
magnitude of each computed cumulant. This does not replace a full convergence analysis,
but it gives a compact check that the plotted cumulants are not dominated by population
accumulating at the edge of the retained Fock space.

\subsection{Benchmarking protocol}
\label{app:benchmarking-protocol}

For both models, the benchmarked operation is the call computing the first two
cumulants after the Hamiltonian, jump operators, monitored jumps, weights, and steady
state have been constructed. The plotted runtimes are arithmetic means over the sample
counts listed in Tables~\ref{tab:linearised-benchmark-params}
and~\ref{tab:dense-benchmark-params}. In \texttt{QuantumFCS.jl} this call is
\begin{center}
    \code{fcscumulants\_recursive(H, J, mJ, 2, rho\_ss, nu)}.
\end{center}
In \texttt{Melt}, the corresponding timing includes calls to \code{FCSAverage} and
\code{FCSNoise} on the already constructed Liouvillian and steady state. The Melt
timings are averaged separately from the Julia timings because high-dimensional
Wolfram evaluations are substantially more expensive. The \texttt{Melt} scripts
export timings in microseconds, which are converted to milliseconds before plotting.

\FloatBarrier
\clearpage

\bibliography{reference}

\end{document}